# Mechanical Analog for Cities

Nicos Makris[1], Gholamreza Moghimi[2], Eric Godat[3], and Tue Vu[4]

## ABSTRACT

Motivated from the increasing need to develop a quantitative, science-based, predictive understanding of the dynamics and response of cities when subjected to hazards, in this paper we apply concepts from statistical mechanics and microrheology to develop mechanical analogs for cities with predictive capabilities. We envision a city to be a matrix where people (cell-phone users) are driven by the city's economy and other associated incentives while using the collection of its infrastructure networks in a similar way that thermally driven Brownian probe particles are moving within a complex viscoelastic material. Mean-square displacements (ensemble averages) of thousands of cell-phone users are computed from GPS location data to establish the creep compliance and the resulting impulse response function of a city. The derivation of these time-response functions allows the synthesis of simple mechanical analogs that model satisfactorily the city's behavior under normal conditions. Our study concentrates on predicting the response of cities to acute shocks (natural hazards that stress the entire urban area) that are approximated with a rectangular pulse with finite duration; and we show that the solid-like mechanical analogs for cities that we derived predict that cities revert immediately to their pre-event response suggesting that they are inherently resilient. Our findings are in remarkable good agreement with the recorded response of the Dallas metroplex following the February 2021 North American winter storm which happened at a time for which we have dependable GPS location data.

## I. INTRODUCTION

Cities are complex systems that traditionally have been generating creativity, growth, leadership, and power together with economic, intellectual, and social wealth. They are complex networks or even networks of complex networks [1] including the transportation infrastructure (roads, bus-lines, and a subway network when available), water utility networks, the electricity grid, telecommunication networks (telephone and internet) together with overriding social networks [2-4]. All these complex networks have been created and upgraded independently during different times with no apparent common design principles. Urbanization is an ever-growing, pressing challenge. As recently as 1950, only 30% of the world's population lived in cities. Today more than 50% is urbanized, and by 2050 more than 3/4 of the world's population is expected to live in cities [5,6]. As cities continue to grow, many of them along the costs of continents which are prone to natural hazards, urban resilience has become a subject of interest in various disciplines [7-10]. The many challenges associated with rapid urban growth, community resilience, and sustainable development are receiving increasing attention; yet, in practice, they have been treated as independent issues [11]. Linear per-capita indicators that conflate distinct dynamic features pertinent to the dynamics specific to each city are typically used to assess these growing challenges; however, it appears these fail to yield direct measures of the impact that hazards can have on cities (such as earthquakes or hurricanes) in association with the need to quantify urban resilience and functional recovery [12].

---

[1] Professor, Dept. of Civil and Environmental Engineering, Southern Methodist Univ, Dallas, TX 75275 (corresponding author). Email: nmakris@smu.edu
[2] Post-Doctoral Fellow, Dept. of Civil and Environmental Engineering, Southern Methodist Univ, Dallas, TX 75275.
[3] Team Lead, Data Science and Research Services, OIT, Southern Methodist Univ, Dallas, TX 75275.
[4] Research Scientist, Data Science and Research Services, OIT, Southern Methodist Univ, Dallas, TX 75275.

Several parallels have been suggested between cities and other complex systems-from biological organisms [13-16] to insect colonies [17] and ecosystems [18]. Cities, while they appear as web-like structures, they are much more than biological organisms or anthills as they involve perpetual exchanges among their citizens who in their majority are productive, creative and innovative individuals, striving for upward mobility and social interaction. Accordingly, while the functionality of cities is served with the superposition of a variety of large-scale, complex networks (transportation, water, electricity, telecommunication) with no apparent common design principles, it appears that cities display a difficult-to-interpret self-organizing behavior that leads to deterministic patterns that can be possibly described with an emerging mechanism yet to be identified [2,6]. Past studies [19,20] have uncovered that, in contrast to the random trajectories predicted by the initially proposed random walk models [21], human trajectories show a high degree of temporal and spatial regularity.

Part of the success of natural sciences hinges upon reductionism where analysis frameworks and methodologies are developed for predicting the emergent macroscopic behavior of a system by monitoring and analyzing the behavior and/or properties of its constituents. For instance, in the emerging field of microrheology, the bulk frequency- and time-response functions of complex viscoelastic materials are inferred by monitoring the thermally driven Brownian motion of probe microspheres immersed within the viscoelastic material and subjected to the perpetual random forces from the collision of the molecules of the viscoelastic material [22-32].

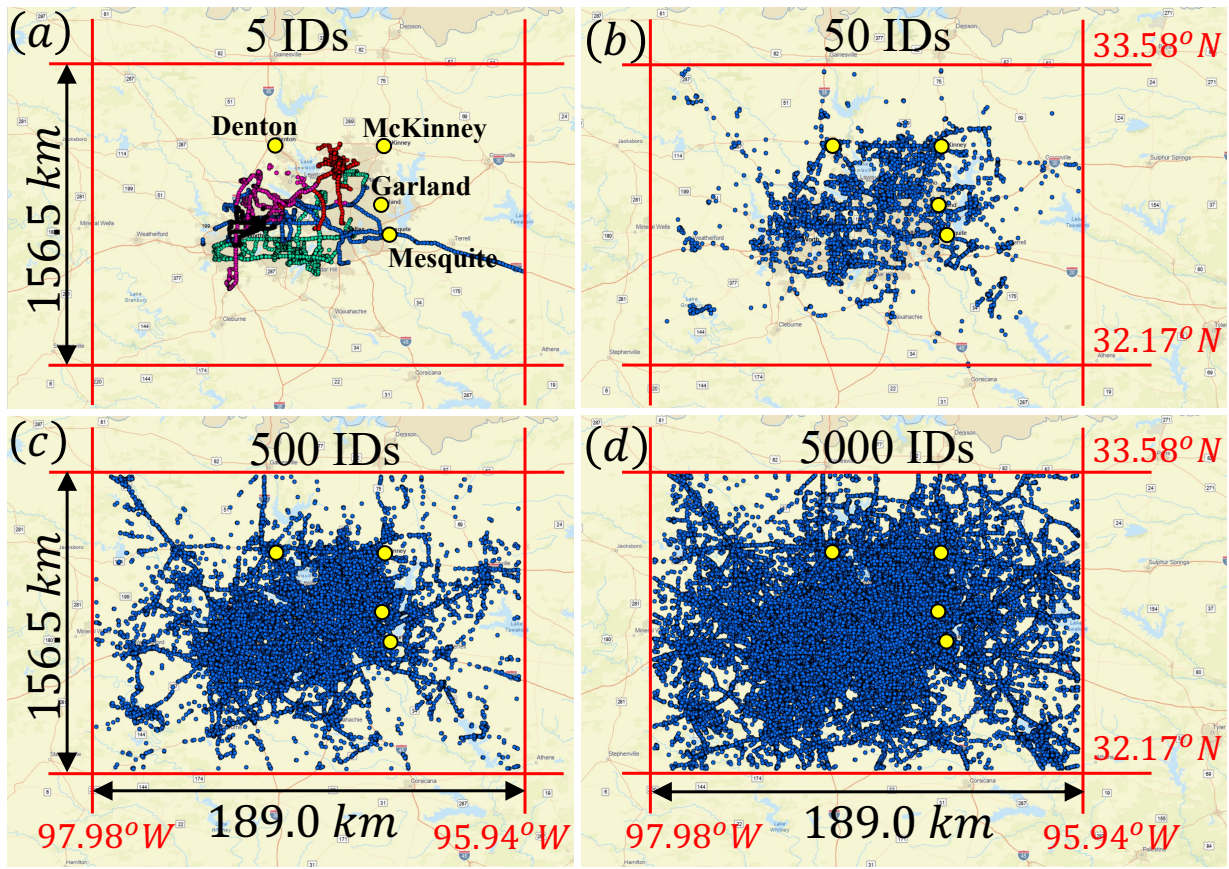

**FIG. 1.** GPS locations of anonymous cell-phone users (IDs) in the greater Dallas metroplex recorded during February and March 2021: (a) 5 IDs; (b) 50 IDs; (c) 500 IDs; (d) 5000 IDs; The yellow dots are nearby cities mentioned in Fig 1(a).

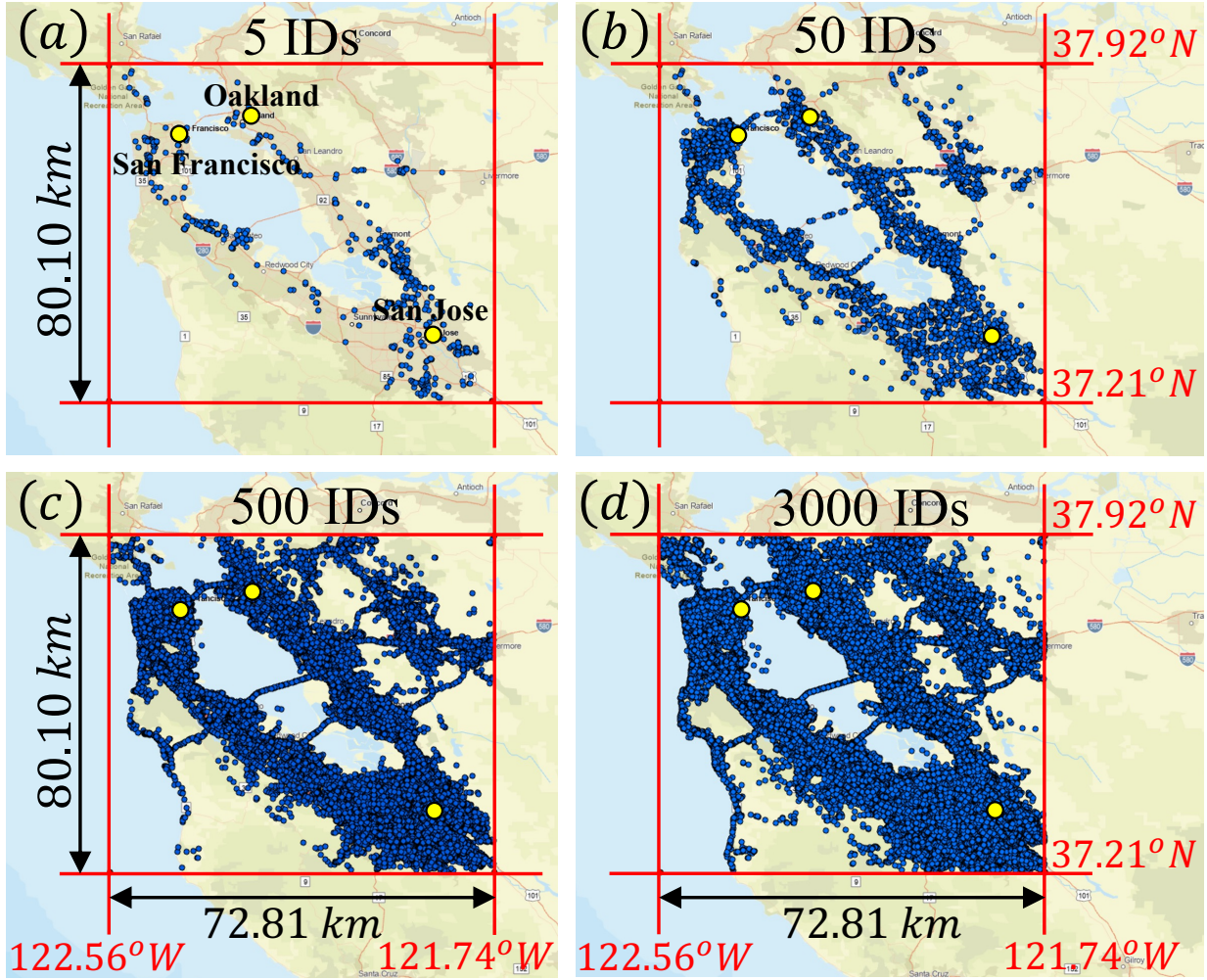

**FIG. 2.** GPS locations of anonymous cell-phone users (IDs) in the San Francisco Bay Area recorded during February and March 2021: (a) 5 IDs; (b) 50 IDs; (c) 500 IDs; (d) 3000 IDs; The yellow dots are nearby cities mentioned in Fig 2(a).

     Building on past studies on human mobility patterns [19,20] which uncovered that citizens exhibit characteristic travel distances in association with a significant probability to return to a few highly frequent locations—a behavior that is reminiscent to particles in a potential well; in this work we envision a city to be a matrix where people (cell-phone users) are driven by the city's economy and other personal and societal driving incentives while using the collection of its infrastructure networks in a similar way that thermally driven Brownian probe particles are moving within a complex viscoelastic material [22-32]. Mobile phones carried by the citizens of an urban center offer rich data on human mobility patterns and we show that the ensemble averages of human GPS location data reveal information that characterizes the dynamics of the entire urban center. Upon computing the mean-square displacements (MSD−ensemble averages) of the recorded time histories of the paths of a large number of citizens of a given city, we propose deterministic mechanical networks with a creep compliance that is proportional to the measured MSD. The response analysis of the proposed solid-like networks offers new insights on the mechanics of urban centers and their emergent inherent resilience within the context of "*engineering resilience*" [33].

Meerow et al. [10] identify several variations of the definition of urban resilience in the Engineering, Business/Finance and Social Science literature. Our study concentrates on a "*single-state equilibrium*" which refers to the capacity of a system to revert to its post-disturbance equilibrium state [33]; therefore, adopts the generic definition of resilience—that is *the capacity of a system to recover its initial state and resume normal activity after a shock* [34-36] and hinges upon the premise that a dependable indicator of the engineering resilience of a city [33] is whether the average mobility pattern of its citizens following an acute shock matches the average mobility pattern before the shock. Our work contributes to a quantitative, science-based, predictive understanding of the dynamics and response of cities to acute shocks [2,6] and concludes that large cities are inherently and invariably resilient.

## II. ENSEMBLE AVERAGES OF GPS LOCATIONS OF CELL-PHONE USERS (IDS)

While the trace of the locations of each individual cell-phone ID shown in Figs. 1 and 2 is unique and every individual has a distinct purpose for generating the recorded trace, the collection of all traces of the anonymous cell-phone users in a given city is treated in this work as a random (stochastic) process. For each of the metroplexes examined in this work we processed longitude and latitude data (in degrees from the Greenwich meridian and from the equator) at various times for tens of thousands of anonymous cell-phone users (IDs). The data were processed and the origin of every ID ("home") was identified by extracting the most frequent occupied location by the ID during the time interval 1:00 AM to 4:00 AM. With reference to Fig. 3, upon the origin="home" of every ID, is established, the East-West (longitude) and North-South (latitude) displacement of each $ID_j$ at time $t_k$ is calculated from [37]

$$x_j(t_k) = \frac{\pi}{180}\left[Long_j(t_k) - LongH_j\right] \cos\left(\frac{\pi}{180} \frac{Lat\, H_j(t_k) + Lat_j(t_k)}{2}\right) R_E \qquad (1)$$

$$y_j(t_k) = \frac{\pi}{180}\left[Lat_j(t_k) - Lat\, H_j\right] R_E \qquad (2)$$

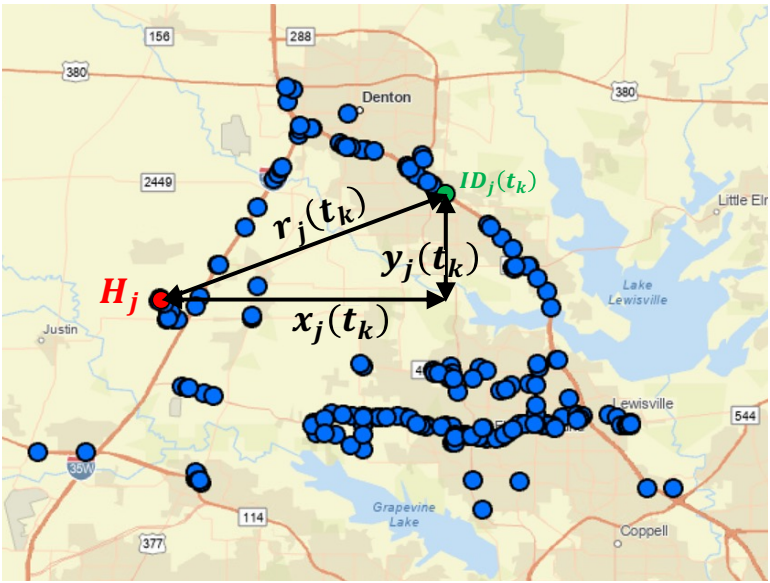

**FIG. 3.** Displacement $r_j(t_k)$ of $ID_j$ at time $t_k$ from its "home" $H_j$. In statistical mechanics and microrheology the symbol $\Delta r_j(t_k)$ rather than $r_j(t_k)$ is often used to express the distance of a probe from its "origin". In this work the symbol $\Delta$ is dropped since the displacement $r_j$ is always measured from the fixed origin $H_j$ at all times.

where $R_E = radius\ of\ Earth \approx 6371\ km$ and the displacement of $ID_j$ at time $t_k$ is $r_j(t_k) = \sqrt{x_j^2(t_k) + y_j^2(t_k)}$ as shown in Fig. 3. For higher precision one can use the *haversine formula*; yet the differences in the result from Eqs. (1) and (2) are negligible given the relative small bounding areas of interest shown in Figs. 1 and 2. From Eqs. (1) and (2) when $ID_j$ moves to the east of its origin (home), $H_j$, its displacement projection, $x_j$ is negative; whereas, when it moves to the west of its home, $H_j$, its displacement projection, $x_j$ is positive. Similarly, when $ID_j$ moves to the north of its home, $H_j$, its displacement projection, $y_j$ is positive; whereas, when it moves to the south of its home, $H_j$, its displacement projection, $y_j$ is negative. Accordingly, for every city the GPS location data of $M$ IDs are organized as shown in Table 1. For every time $t_k$ appearing in the first column of Table I (say top of the hour), the location $x_j(t_k), y_j(t_k)$ is the GPS location of $ID_j$ recorded closer to that time $t_k$.

**TABLE I.** GPS location data of $M$ cell-phone users (IDs) recorded closest to the top of the hour.

| Time | $ID_1$ | | | ... | $ID_j$ | | | ... | $ID_M$ | | |
|---|---|---|---|---|---|---|---|---|---|---|---|
| $t_1$ | $x_1(t_1)$ | $y_1(t_1)$ | $r_1(t_1)$ | ... | $x_j(t_1)$ | $y_j(t_1)$ | $r_j(t_1)$ | ... | $x_M(t_1)$ | $y_M(t_1)$ | $r_M(t_1)$ |
| $t_2$ | $x_1(t_2)$ | $y_1(t_2)$ | $r_1(t_2)$ | ... | $x_j(t_2)$ | $y_j(t_2)$ | $r_j(t_2)$ | ... | $x_M(t_2)$ | $y_M(t_2)$ | $r_M(t_2)$ |
| ⋮ | ⋮ | ⋮ | ⋮ | ... | ⋮ | ⋮ | ⋮ | ... | ⋮ | ⋮ | ⋮ |
| $t_k$ | $x_1(t_k)$ | $y_1(t_k)$ | $r_1(t_k)$ | ... | $x_j(t_k)$ | $y_j(t_k)$ | $r_j(t_k)$ | ... | $x_M(t_k)$ | $y_M(t_k)$ | $r_M(t_k)$ |
| ⋮ | ⋮ | ⋮ | ⋮ | ... | ⋮ | ⋮ | ⋮ | ... | ⋮ | ⋮ | ⋮ |
| $t_n$ | $x_1(t_n)$ | $y_1(t_n)$ | $r_1(t_n)$ | ... | $x_j(t_n)$ | $y_j(t_n)$ | $r_j(t_n)$ | ... | $x_M(t_n)$ | $y_M(t_n)$ | $r_M(t_n)$ |

Figure 4 plots time-histories of the east-west $(E-W)$ movement, $x_j(t), j \in \{1,2,...,M\}$ of selected IDs from the Dallas metroplex. Some IDs move frequently both to the east and to the west, other IDs move systematically to the east, other IDs move systematically to the west due to their daily routine (say travelling to their workplace). Accordingly, the time-averages from individual time-histories may differ drastically, therefore, the stochastic (random) process shown in Fig. 4 is clearly non-ergodic [38,39]. With reference to Fig. 4 for every time $t_k$ we compute ensemble averages of the east-west $(E-W)$ movement $\langle x(t_k) \rangle$ and the north-south $(N-S)$ movement $\langle y(t_k) \rangle$ (mean values) for all the $M$ available IDs

$$\langle x(t_k) \rangle = \frac{1}{M} \sum_{j=1}^{M} x_j(t_k) \quad \text{and} \quad \langle y(t_k) \rangle = \frac{1}{M} \sum_{j=1}^{M} y_j(t_k) \qquad (3)$$

Fig. 5 plots the mean values (ensemble averages) $\langle x(t_k) \rangle$ (left) and $\langle y(t_k) \rangle$ (right) of $M = 13,000$ IDs in the Dallas metroplex for all times $t_k$ from Feb 01 to March 31, 2021. Fig. 5 uncovers that the time-histories of the mean values $\langle x(t) \rangle$ and $\langle y(t) \rangle$ fluctuate within only a couple of hundred meters.

Similarly, for every time $t_k$ and a given time shift, $\tau$, we compute the location (position) autocorrelation functions of the east-west $(E-W)$ movement $<x(t_k).x(t_k+\tau)>$ and the north-south $(N-S)$ movement $\langle y(t_k).y(t_k+\tau) \rangle$ for all the $M$ available IDs in a city (ensemble).

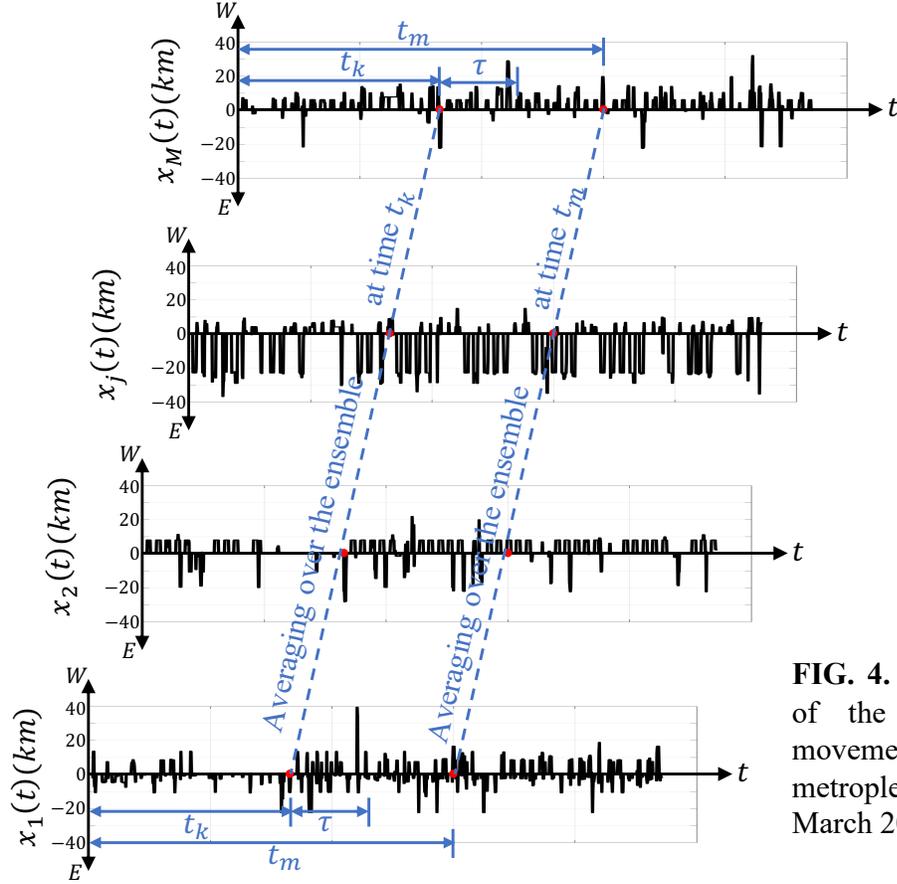

**FIG. 4.** Ensemble time histories of the longitudinal $(E-W)$ movement of IDs from the Dallas metroplex during February and March 2021.

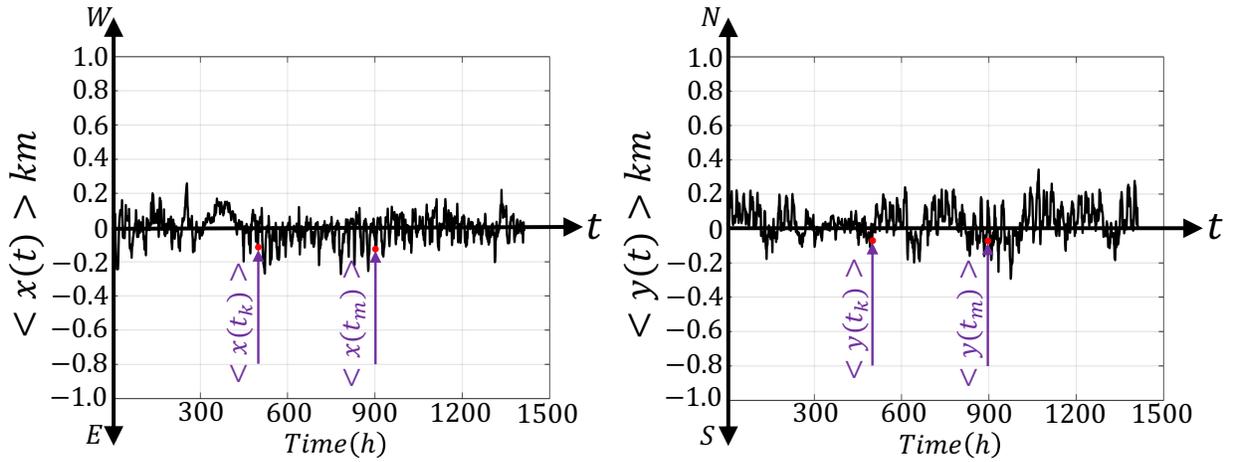

**FIG. 5.** Time-histories of the mean values (ensemble averages) of the $E-W$ movement $\langle x(t_k) \rangle$ (left) and $N-S$ movement $\langle y(t_k) \rangle$ (right) of $M = 13{,}000$ IDs in the Dallas metroplex during February and March 2021.

$$< x(t_k).x(t_k + \tau) > = \frac{1}{M} \sum_{j=1}^{M} x_j(t_k) x_j(t_k + \tau) \tag{4a}$$

and

$$< y(t_k).y(t_k + \tau) > = \frac{1}{M} \sum_{j=1}^{M} y_j(t_k) y_j(t_k + \tau) \tag{4b}$$

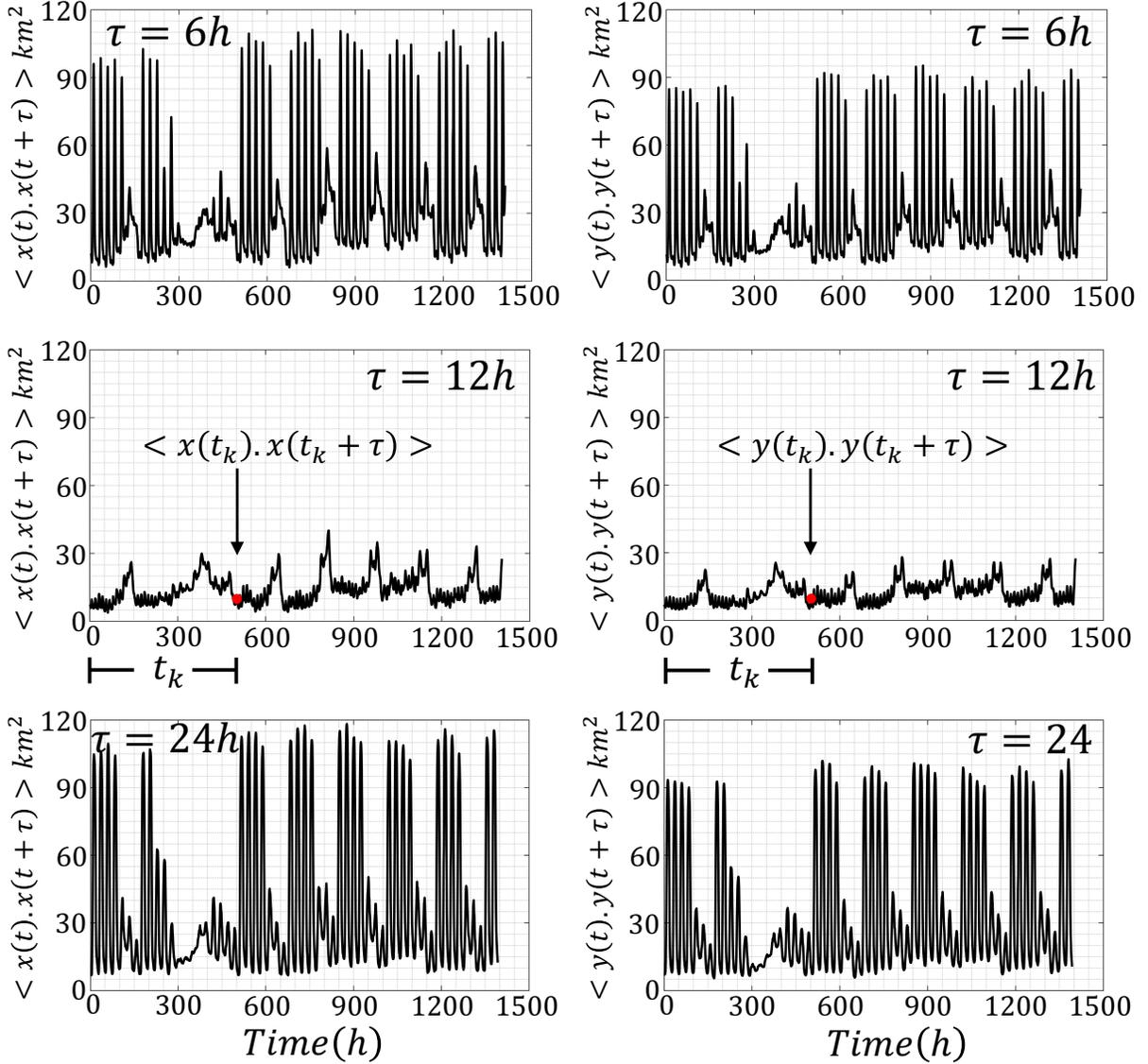

**FIG. 6.** Time-histories of the location (position) autocorrelation functions (ensemble averages) of the $E-W$ movement $< x(t_k).x(t_k + \tau) >$ (left) and the $N-S$ movement $< y(t_k).y(t_k + \tau) >$ (right) of the $M = 13,000$ IDs in the Dallas metroplex during February and March 2021 for three different values of the time shift $\tau = 6h$, $12h$ and $24h$.

Fig. 6 plots the location (position) autocorrelation functions (ensemble averages) $<x(t_k).x(t_k+\tau)>$ (left) and $<y(t_k).y(t_k+\tau)>$ (right) of $M=13,000$ IDs in the Dallas metroplex for all times $t_k$ and time shifts $\tau=6h, 12h$ and $24h$. Clearly, for $\tau=24h$, the location autocorrelation functions shown in Fig. 6 show a stronger correlation than when $\tau=12h$, since after 24 hours the cell-phone users are likely to be at the same location that they were the day before. Accordingly, while the mean values of the locations (positions) shown in Fig. 5 fluctuate within only a couple of hundred meters, the strong fluctuations over time of the position autocorrelation functions shown in Fig. 6 indicate that the stochastic (random) process schematically illustrated in Fig. 4 does not even satisfy the notion of a weakly stationary process [38,39]. Accordingly, in this study we work with ensemble averages (Not time averages).

### III. THEORY

The analysis framework that we develop herein to construct a mechanical analog for cities is inspired from major advances in microrheology in the mid-to-late 1990s [22-32] which make possible the extraction of the bulk mechanical properties of complex viscoelastic materials by monitoring the thermally driven Brownian motion of probe microspheres suspended within the viscoelastic material.

The phenomenon of Brownian motion was first explained in the 1905 Einstein's celebrated paper [40] which examined the long-term response of Brownian microspheres with mass $m$ and radius $R$ suspended in a memoryless Newtonian fluid with viscosity $\eta$. Einstein's theory of Brownian motion predicts the long-term expression of the mean-square displacement, $\langle r^2(t)\rangle = MSD$, of the randomly moving microspheres (diffusive regime)

$$MSD = \langle r^2(t)\rangle = \frac{1}{M}\sum_{j=1}^{M} r_j^2(t) = 2NDt = \frac{NK_BT}{3\pi R}\frac{t}{\eta} \qquad (5)$$

where $M$ is the number of suspended microspheres, while, $r_j(t)$ is the distance of microsphere $j$ at time $t$ from some origin. In Eq. (5) $N \in \{1,2,3\}$ is the number of spatial dimensions, $K_B$ is Boltzmann's constant, $T$ is the equilibrium temperature of the Newtonian fluid with viscosity $\eta$ within which the $M$ Brownian microspheres are immersed and $D = K_BT/6\pi R\eta$ is the time-independent diffusion coefficient. Eq. (5) derived by Einstein in 1905, in association with Stokes' result—that the drag force on a slowly moving sphere with velocity $v$, is $6\pi R\eta v$ [41], is of central interest in statistical mechanics since it relates the thermally driven ensemble average $\langle r^2(t)\rangle$ from a stochastic process to the deterministic emergent property of the Newtonian fluid—that is its viscosity, $\eta$. The time-derivative of Eq. (5) gives

$$\frac{d\langle r^2(t)\rangle}{dt} = \frac{2}{M}\sum_{j=1}^{M}\left|r_j(t)\frac{dr_j(t)}{dt}\right| = 2ND = \frac{NK_BT}{3\pi R\eta} \qquad (6)$$

The reader recognizes that the right-hand-side term of Eq. (6) is a constant $= NK_BT/3\pi R\eta = 2ND$; whereas, the left-hand side $\frac{2}{M}\sum_{j=1}^{M}\left|r_j(t)\frac{dr_j(t)}{dt}\right|$ is zero when $t=0$, since $r_j(0)=0 \quad \forall j \in \{1,2,...,M\}$. This inconsistency emerges because Eq. (5) is valid only in the long-term (diffusive regime at large times). At short-time scales, when $t < m/6\pi R\eta = \tau$, the Brownian motion of suspended particles is influenced by the inertia of the particle and the surrounding fluid (ballistic regime) and Einstein's 1905 "long-term" result offered by Eq. (5) was extended for all time-scales by Uhlenbeck and Ornstein [42].

$$\langle r^2(t) \rangle = \frac{NK_BT}{3\pi R} \frac{1}{\eta} \left[ t - \tau\left(1 - e^{-t/\tau}\right) \right] \tag{7}$$

Equation (7) yields that at $t = 0$, $d\langle r^2(t)\rangle/dt = 0$ which is in agreement with the left-hand side of Eq. (6) when $t = 0$.

### A. Generalization for Brownian motion within non-newtonian fluids

In their seminal paper, Mason and Weitz [22] employed dynamic light scattering to measure the mean-square displacement $\langle r^2(t) \rangle$ of probe Brownian particles immersed in a linear viscoelastic fluid and related it to the complex dynamic modulus, $\mathcal{G}_{ve}(s)$ of the viscoelastic fluid within which the probe particles are immersed. Given the viscoelastic behavior of the complex fluid, the motion of a probe Brownian particle is described by the generalized Langevin equation [43,44]

$$m \frac{dv(t)}{dt} = -\int_{0^-}^{t} \zeta(t - \xi) v(\xi) d\xi + f_R(t) \tag{8}$$

where $m$ is the mass of the Brownian particle, $v(t)$ is the particle velocity and $f_R(t)$ are the random forces acting on the particle from the collisions of the fluid modulus on the Brownian particle. The integral in Eq. (8) represents the drag force on the particle as it moves through the viscoelastic fluid and accounts for the fading memory of this drag due to the elasticity of the viscoelastic fluid. Upon transforming Eq. (8) in the Laplace domain, Mason and Weitz [22] reached the following result for the Laplace transform of the mean-square displacement

$$\langle r^2(s) \rangle = \int_0^\infty \langle r^2(t) \rangle e^{-st} dt = \frac{NK_BT}{3\pi R} \frac{1}{s\left(\mathcal{G}_{ve}(s) + \frac{m}{6\pi R}s^2\right)} \tag{9}$$

where $s$ is the Laplace variable. In deriving Eq. (9) a key assumption was adopted—that the Stokes drag coefficient on a sphere moving slowly in a memoryless, Newtonian viscous fluid, $\zeta = 6\pi R \eta$ [41], can be generalized to relate the impedance of the Brownian particle, $\mathcal{Z}(s) = \mathcal{L}\{\zeta(t)\} = \int_{0^-}^\infty \zeta(t) e^{-st} dt$ to the complex dynamic viscosity of the viscoelastic material, $\eta_{ve}(s) = \mathcal{G}_{ve}(s)/s$. Accordingly, Mason and Weitz [22] made the physically motivated assumption that

$$\mathcal{Z}(s) = \int_0^\infty \zeta(t) e^{-st} dt = 6\pi R \eta_{ve}(s) = 6\pi R \frac{\mathcal{G}_{ve}(s)}{s} \tag{10}$$

In a recent publication, Makris [45] showed that the time response function $J(t) = \frac{1}{\eta}\left[t - \tau(1 - e^{-t/\tau})\right]$ appearing on the right-hand side of Eq. (7) is the creep compliance (retardation function) of a linear network where a dashpot with viscosity $\eta$ is connected in parallel with an inerter with distributed inertance $m_R = m/6\pi R$. Building on the work of [42,46] in association with the work of Mason and Weitz [22], Makris [45] uncovered a viscous-viscoelastic correspondence principle for Brownian motion which states that the mean-square displacement $\langle r^2(t) \rangle$ of Brownian particles (microspheres) with mass $m$ and radius $R$ suspended in some linear, isotropic viscoelastic material when subjected to the random forces from the collisions of the molecules of the viscoelastic material, is

$$\langle r^2(t) \rangle = \frac{NK_BT}{3\pi R} J(t) \tag{11}$$

where $J(t)$ is the creep compliance (strain-history due to a unit-step stress) of a viscoelastic network that is a parallel connection of the linear viscoelastic material within which the Brownian particles are immersed and an inerter with distributed inertance $m_R = m/6\pi R$. Equation (11) also holds for the case where the density of the fluid surrounding the Brownian microparticles is appreciable [47] and in this case in addition to the Stokes viscous drag, the Brownian motion of the immersed microparticles develops the Boussinesq hydrodynamic memory [48].

Equation (11), which is the equivalent of Eq. (9) in the time domain, generalizes the main result of Eq. (5) and relates the thermally driven ensemble average $\langle r^2(t) \rangle = MSD$ from a stochastic (random) process to the deterministic creep compliance (retardation function), $J(t)$ of a mechanical network. In this work we build upon Eq. (11) in an effort to propose a corresponding relation for cities,

$$\langle r^2(t) \rangle = WJ(t) \tag{12}$$

where $\langle r^2(t) \rangle$ is the mean-square displacement of cell-phone users as computed by averaging over a city ensemble of GPS location traces, $J(t)$ is the creep compliance (displacement-history due to a unit-step force) of a mechanical model to be identified and $W$ is a proportionality constant having the units of energy (force×length).

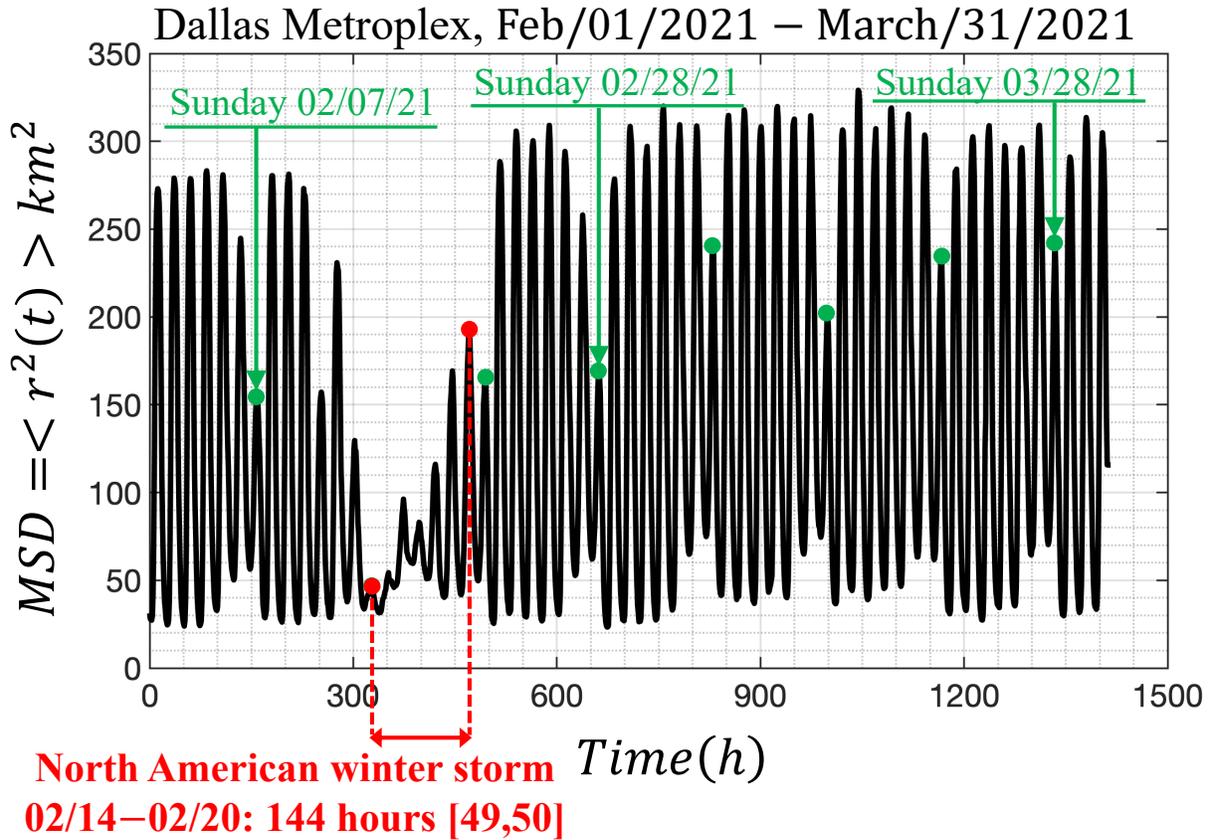

**FIG. 7.** Time-history of the mean-square displacement $= \langle r^2(t) \rangle = \frac{1}{M}\sum_{j=1}^{M} r_j^2(t)$ (ensemble average) of $M = 13{,}000$ IDs in the Dallas metroplex during February and March 2021. Dots (in green) atop the suppressed spikes indicate Sundays.

The motion of people in a city as recorded with their GPS locations happens in a matterless environment (the environment of a city which does not exhibit mechanical elasticity, dissipation or inertia); while it is driven by conceptual incentives; therefore, we do not have a Langevin equation to describe the motion of people in cities as we have Eq. (8) that describes the random motion of Brownian particles immersed in some linear, isotropic viscoelastic material. Furthermore, the statistical analysis of the GPS location data processed in this study yields that ensemble averages as illustrated in Fig. 4 do not even support the notion of a weak stationary random process [38,39]. Nevertheless, in view of the overall phenomenological similarities between the motion of people in a city (see Figs. 1 and 2) and the motion of probe Brownian microparticles in a viscoelastic material in association with the remarkable success of Eqs. (9) or (11) in microrheology, we adopt Eq. (12) in an effort to develop a mechanical model for cities with potential engineering significance. Fig. 7 plots the time-history of the mean-square displacement $(MSD) = \langle r^2(t) \rangle$

$$\langle r^2(t) \rangle = \frac{1}{M}\sum_{j=1}^{M} r_j^2(t) = \frac{1}{M}\sum_{j=1}^{M} x_j^2(t) + y_j^2(t) \qquad (13)$$

of $M = 13,000$ cell-phone users (IDs) in the Dallas metroplex during February and March 2021; while, Fig. 8 plots the time-history of $\langle r^2(t) \rangle$ of $M = 3,400$ cell-phone users (IDs) in the San Francisco Bay Area during the same time.

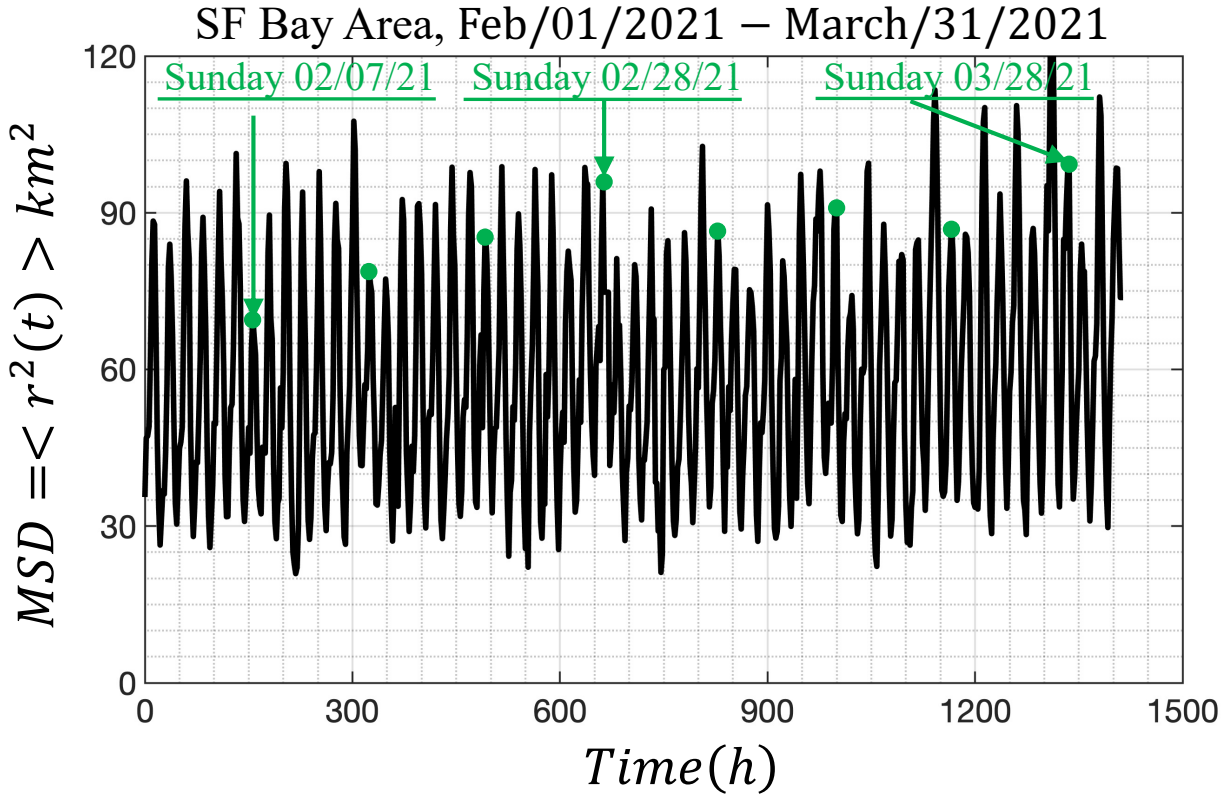

**FIG. 8.** Time-history of the mean-square displacement $= \langle r^2(t) \rangle = \frac{1}{M}\sum_{j=1}^{M} r_j^2(t)$ (ensemble average) of $M = 3,400$ IDs in the San Francisco Bay Area during February and March 2021. Dots (in green) atop the suppressed spikes indicate Sundays.

Figure 7 reveals that the MSD for the Dallas metroplex exhibits a remarkable periodicity with people moving more during the weekdays and less on Sundays which are marked with green dots. At the same time, less people are staying at home during the weekends than during the weekdays. The MSD from the Dallas metroplex exhibits a notable suppression during the major North American winter storm that happened during the $3^{rd}$ week of February 2021 and ended sometime towards the end of that week before Sunday, Feb 21, 2021 [49,50]. Over 4 million people lost power due to the storm which resulted to a state-wide power crisis in Texas causing major damage to the Texas power grid. Regardless of its unprecedented intensity and resulting disruption from the winter storm; Fig 7 reveals that immediately after the cold winter storm, the city of Dallas resumes normal activity with the MSD of its citizens to revert to its normal oscillation pattern as recorded before the storm, suggesting that the Dallas metroplex exhibits a great degree of engineering resilience [33].

The MSD from the San Francisco Bay Area for the same period (Feb 01−March 31, 2021) is less ordered that the MSD from the Dallas metroplex; nevertheless the same pattern of less people staying at home during the weekends than during the weekdays is observed. Furthermore, the average MSD from the SF Bay Area covers less area (on an average $60\ km^2$) when compared to $165\ km^2$ from the Dallas metroplex. Both cities "never sleep" since the MSDs from both the Dallas metroplex (Fig 7) and from the San Francisco Bay Area (Fig. 8) reach a bottom at approximately $30\ km^2$.

## IV.  CANDIDATE MECHANICAL ANALOGS FOR CITIES

Figures 7 and 8 show that the mean-square displacement, $\langle r^2(t) \rangle$, of cell-phone users as computed by averaging over an ensemble of citizens from major urban areas, is essentially an elevated cosine with a daily period where the amplitude of the cosine oscillation is modulated with a weekly period. Accordingly in view of Eq. (12), in this section we are in search of mechanical networks that their creep compliance, $J(t)$ (displacement-history due to a unit-step force) can approximate the time-history of the mean-square displacement, $\langle r^2(t) \rangle$ as computed from ensemble averages and shown in Figs. 7 and 8.

Given that the mean-square displacements, $\langle r^2(t) \rangle$, shown in Figs. 7 and 8 are bounded functions, in association with Eq. (12) we are in search of a solid-like network so that its creep compliance, $J(t)$ is a bounded time response function. Furthermore, the perpetual undamped oscillations of the computed $\langle r^2(t) \rangle$ with no visible rise-time imply that the candidate mechanical network shall be undamped. Finally, given that the cosine function shown in Figs. 7 and 8 is elevated above the horizontal axis, the candidate network needs to yield a creep compliance function that is more elaborate than the elementary inertoelastic solid (the spring-inerter parallel connection) [51,52].

### A.  The three-parameter inertoelastic solid

The simplest mechanical network that meets the above requirements is a parallel connection of an elastic spring with an inertoelastic fluid (spring-inerter in-series connection) as shown in Fig. 9. The total force $F(t) = F_1(t) + F_2(t)$ from the linear network shown in Fig. 9 is the summation of the force output from linear spring with elastic constant $k_1$,

$$F_1(t) = k_1\, u(t) \qquad (14)$$

and the force output from the inertoelastic fluid [51]

$$F_2(t) + \frac{1}{\omega_{R2}^2} \frac{d^2 F_2(t)}{dt^2} = M_R \frac{d^2 u(t)}{dt^2} \tag{15}$$

where $M_R$ is the inertance of the inerter with units of mass [M] and $\omega_{R2}^2 = k_2/M_R$ is the rotational frequency of the inertoelastic fluid. The summation of Eqs. (14) and (15) together with the second time-derivative of Eq. (14) yields the constitutive equation of the three-parameter inertoelastic solid shown in Fig. 9.

$$F(t) + \frac{1}{\omega_{R2}^2} \frac{d^2 F_2(t)}{dt^2} = k_1 \left[ u(t) + \left( \frac{1}{\omega_{R1}^2} + \frac{1}{\omega_{R2}^2} \right) \frac{d^2 u(t)}{dt^2} \right] \tag{16}$$

where $\omega_{R1}^2 = k_1/M_R$. Upon replacing

$$\frac{1}{\omega_{R1}^2} + \frac{1}{\omega_{R2}^2} = \frac{1}{\omega_R^2} \implies \omega_R^2 = \frac{\omega_{R1}^2 \omega_{R2}^2}{\omega_{R1}^2 + \omega_{R2}^2} = \frac{\omega_{R2}^2}{1 + \frac{\omega_{R2}^2}{\omega_{R1}^2}}, \tag{17}$$

the Laplace transform of Eq. (16) gives $u(s) = \mathcal{L}\{u(t)\} = \int_0^\infty u(t) e^{-st} dt = \mathcal{H}(s) F(s)$, where $\mathcal{H}(s)$ is the complex dynamic flexibility of the linear network shown in Fig. 9.

$$\mathcal{H}(s) = \frac{u(s)}{F(s)} = \frac{1}{k_1} \frac{1 + \frac{s^2}{\omega_{R2}^2}}{1 + \frac{s^2}{\omega_R^2}} = \frac{1}{k_1} \left[ \frac{\omega_R^2}{s^2 + \omega_R^2} + \frac{\omega_R^2}{\omega_{R2}^2} \frac{s^2}{s^2 + \omega_R^2} \right] \tag{18}$$

The Laplace transform of the creep compliance, $J(t)$ is the complex creep function $\mathcal{C}(s) = \mathcal{L}\{J(t)\} = \int_0^\infty J(t) e^{-st} dt = \mathcal{H}(s)/s$ [53-56]. Upon dividing Eq. (18) with the Laplace variable $s$ we obtain

$$\mathcal{C}(s) = \frac{\mathcal{H}(s)}{s} = \frac{1}{k_1} \left[ \frac{1}{s} \frac{\omega_R^2}{s^2 + \omega_R^2} + \frac{\omega_R^2}{\omega_{R2}^2} \frac{s}{s^2 + \omega_R^2} \right] = \frac{1}{k_1} \left[ \frac{1}{s} - \frac{s}{s^2 + \omega_R^2} + \frac{\omega_R^2}{\omega_{R2}^2} \frac{s}{s^2 + \omega_R^2} \right] \tag{19}$$

The inverse Laplace transform of Eq. (19) offers the creep compliance of the three-parameter inertoelastic solid shown in Fig. 9.

$$J(t) = \mathcal{L}^{-1}\{\mathcal{C}(s)\} = \frac{1}{k_1} \left[ U(t - 0) - \frac{\beta}{1 + \beta} \cos(\omega_R t) \right] \tag{20}$$

where $U(t - 0)$ is the Heaviside unit-step function at the time origin [57] and parameter $\beta = \omega_{R2}^2/\omega_{R1}^2 = k_2/k_1$.

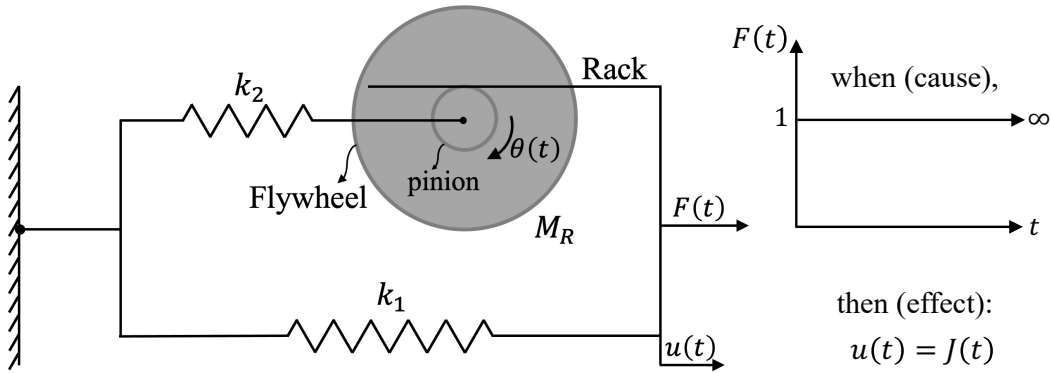

**FIG. 9.** The three-parameter introelastic solid which is a parallel connection of an introelastic fluid (spring-inerter in-series connection) with an elastic spring with stiffness $k_1$.

The expression of the creep compliance, $J(t)$ offered by Eq. (20) is an elevated cosine function with frequency $\omega_R = \omega_{R2}/\sqrt{1+\beta} = 2\pi/24h = \pi/12\ rad/h$. Substitution of Eq. (20) into Eq. (12), which is the central equation for this study, the elementary 3-parameter inertoelastic solid mechanical analog predicts that

$$<r^2(t)> WJ(t) = \frac{W}{k_1}\left[U(t-0) - \frac{\beta}{1+\beta}\cos(\omega_R t)\right] \quad (21)$$

where the proportionality constant $W/k_1 = A_0$ has units of area ($km^2$). Accordingly, given that $\omega_R = \omega_{R2}/\sqrt{1+\beta}$ is the daily frequency ($\omega_R = \pi/12\ rad/h$), only two parameters of the three-parameter inertoelastic solid need to be calibrated from the recorded GPS location data; $A_0 = W/k_1$, and $\beta = k_2/k_1 = \omega_{R2}^2/\omega_{R_1}^2$.

Figures 10 and 11 compare the predictions of the calibrated three-parameter inertoelastic solid shown in Fig. 9 against the recorded $MSD = <r^2(t)>$ of $M = 13{,}000$ IDs in the Dallas metroplex (Fig. 10) and the resulting $MSD = <r^2(t)>$ of 3,400 IDs in the San Francisco Bay Area (Fig. 11) during February and March 2021. The mean-square displacement, $\langle r^2(t)\rangle$, shown in Figs. 10 and 11 as computed by averaging over a city ensemble of GPS location traces suggests that the coefficient $\beta/(1+\beta)$ of the cosine function in Eq. (20) or (21) is the average normalized amplitude of the daily oscillations of $\langle r^2(t)\rangle$. With reference to Fig. 10 for the Dallas metroplex, $\beta/(1+\beta) = (1/2)((300-30)/165) \approx 0.8181 \Rightarrow \beta = 4.5$ with $A_0 = W/k_1 = 165\ km^2$; whereas for the San Francisco Bay Area $\beta/(1+\beta) = (1/2)((90-30)/60) = 0.5 \Rightarrow \beta = 1.0$ with $A_0 = W/k_1 = 60\ km^2$.

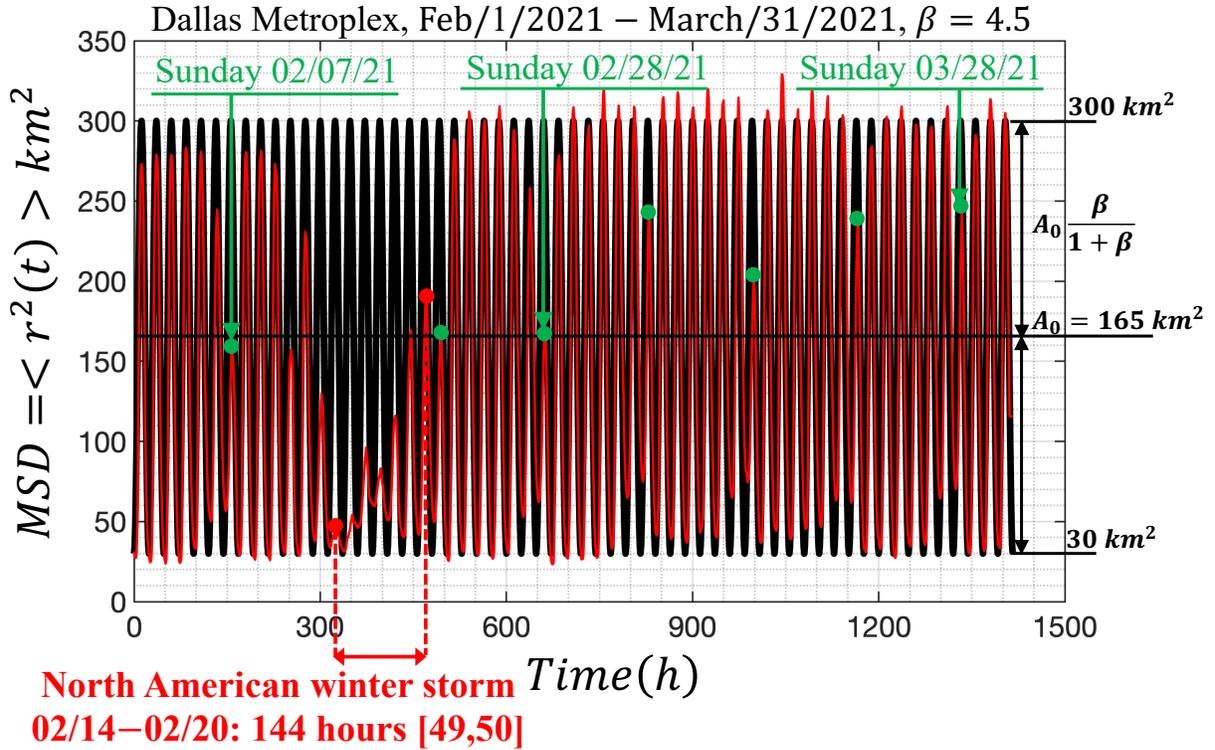

**FIG. 10.** Comparison of the result of Eq. (21), $\langle r^2(t)\rangle = WJ(t)$ with the recorded mean-square displacement, $\langle r^2(t)\rangle$ from the Dallas metroplex during February and March 2021.

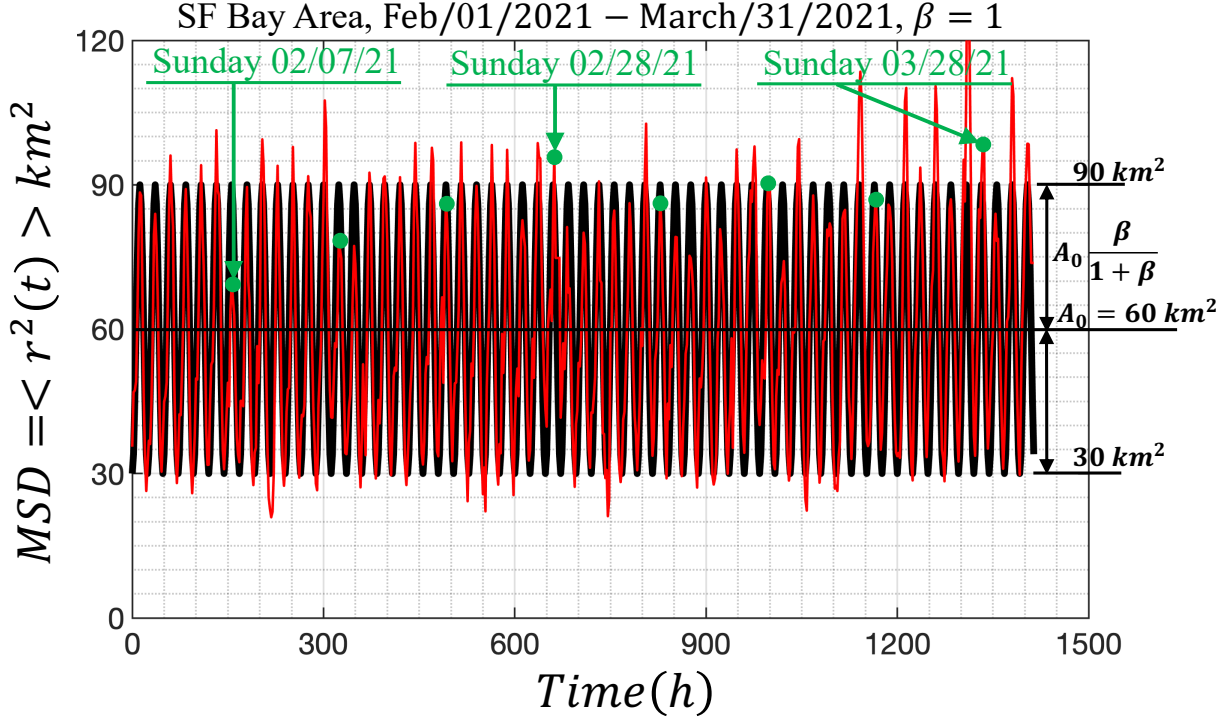

**FIG. 11.** Comparison of the result of Eq. (21), $\langle r^2(t) \rangle = WJ(t)$ with the recorded mean-square displacement, $\langle r^2(t) \rangle$ from the San Francisco Bay Area during February and March 2021.

### B. An amplitude-modulation model

The time-history of the mean-square displacement $\langle r^2(t) \rangle$ of the Dallas metroplex IDs shown in Fig. 7 exhibits a distinguishable weekly modulation (message signal) in addition to the daily oscillations (carrier wave), $C(t) = C \cos(\omega_R t)$ with $\omega_R = 2\pi/24h = \pi/12 \; rad/h$. By expressing the weekly modulating message signal [58,59].

$$s(t) = \gamma \, C \sin\left(\frac{\omega_R}{7} t\right) \quad (22)$$

an improved mathematical expression to Eq. (21) that better approximates the Dallas metroplex mean-square displacement shown in Fig. 7 is

$$\langle r^2(t) \rangle = A_0 \left\{ U(t-0) - C \cos(\omega_R t) \left[1 + \gamma \sin\left(\frac{\omega_R}{7} t\right)\right]\right\} \quad (23)$$

where $C$ and $\gamma$ are dimensionless model parameters.

Upon using the trigonometric identity $\cos(\omega_R t) \sin\left(\frac{\omega_R}{7} t\right) = \frac{1}{2}\left[\sin\left(\frac{8}{7}\omega_R t\right) - \sin\left(\frac{6}{7}\omega_R t\right)\right]$, Eq. (24) assumes the form

$$\langle r^2(t) \rangle = WJ(t) = A_0 \left\{ U(t-0) - C \left[\cos(\omega_R t) + \frac{\gamma}{2}\left(\sin\left(\frac{8}{7}\omega_R t\right) - \sin\left(\frac{6}{7}\omega_R t\right)\right)\right]\right\} \quad (24)$$

where $A_0 = W/k$ is a proportionality constant with units $[L]^2$ that relates the mean-square displacement to a normalized creep compliance (see also Eq. (21)). Parameter $k$ has units of stiffness (force/length). For the mathematical model offered by Eq. (24), we do not have a mechanical model that consists of a collection of spring and inerters as is the three-parameter inertoelastic model shown in Fig. 9. The model expressed by Eqs. (23) or (24) was merely constructed from experience in manipulating signals in radio communications [58,59]. Figures (12) and (13) compare the result of Eq. (24) against the recorded $MSD = \langle r^2(t) \rangle$ in the Dallas metroplex (Fig 12) and the recorded $MSD = \langle r^2(t) \rangle$ in the San Francisco Bay Area (Fig 13) during February and March 2021. The amplitude-modulation expression for $\langle r^2(t) \rangle$ given by Eq. (24) captures the observed behavior that more activity happens during the weekdays, while less people are staying at home during the weekends than during the weekdays.

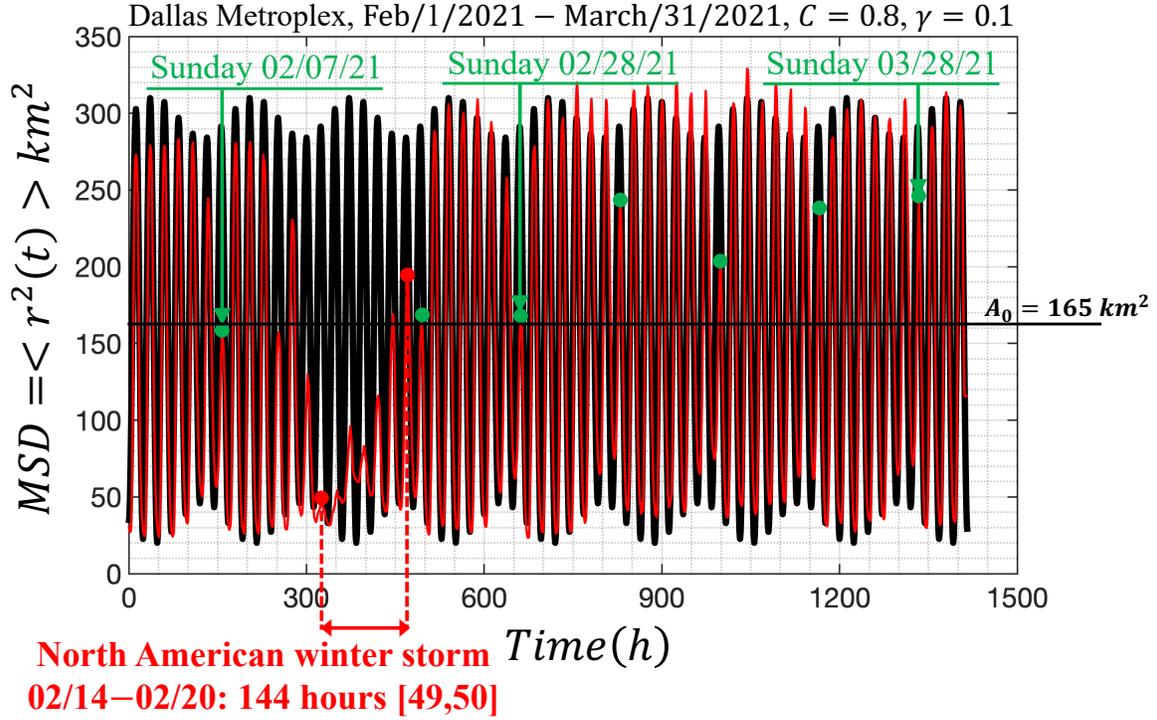

**FIG. 12.** Comparison of the amplitude-modulation expression given by Eq. (24) with the recorded mean-square displacement $\langle r^2(t) \rangle$ of $M = 13,000$ IDs from the Dallas metroplex during February and March 2021.

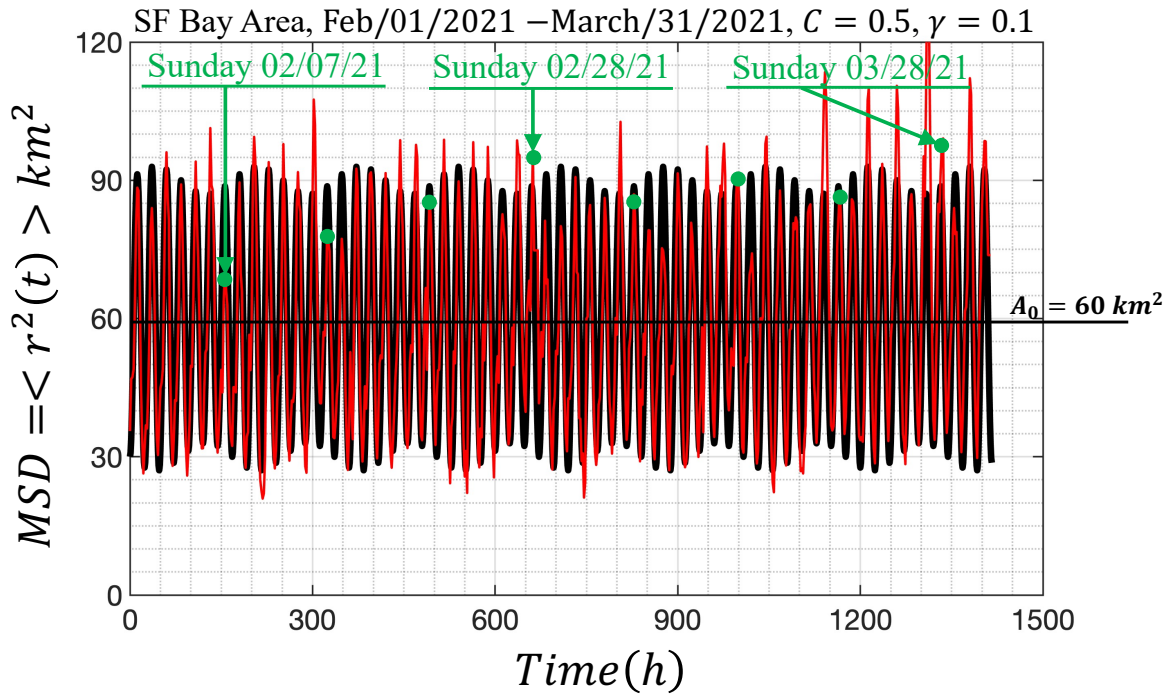

**FIG. 13.** Comparison of the amplitude-modulation expression given by Eq. (24) with the recorded mean-square displacement $\langle r^2(t) \rangle$ of $M = 3,400$ IDs from the San Francisco Bay Area during February and March 2021.

## V. RESPONSE OF THE CANDIDATE MODELS TO A RECTANGULAR PULSE WITH DURATION $T_e$

We return now to the mechanical model depicted in Fig. 9 which is the simplest mechanical analog for a city since its creep compliance $J(t)$ given by Eq. (20) offers according to Eq. (12) a mean-square displacement $\langle r^2(t) \rangle$ that matches to a satisfactory extent the recorded mean-square displacement as results from the GPS location data of several thousand users (see Figs. 10 and 11).

The main motivation of this study is to develop quantitative models for cities that have some predictive capability. Our response analysis starts with an elementary rectangular force pulse of amplitude $F_o$ and duration $T_e$ which is expressed as

$$F(t) = F_o[U(t-0) - U(t-T_e)] \quad (25)$$

where $U(t-\xi)$ is the unit-amplitude Heaviside step function that initiates at time $t = \xi$ [57]. The rectangular pulse with duration $T_e$ expressed by Eq. (25) is suitable to express shocks from natural hazards (earthquakes, hurricanes and cold storms) which start and end abruptly and are system-wide [60]. In contrast, Gaussian functions with long tails are less suitable to approximate natural hazards which impact cities suddenly with little or zero loading build-up time. Pandemics and economic crises are stressing cities during larger time-scales during which the city may develop mechanisms to readjust [10]. Therefore, for such prolonged crisis, different loading patterns than the rectangular pulse described with Eq.(25) may be more suitable. The response of the three-parameter inertoelastic model shown in Fig. 9 which is described with the constitutive Eq. (16) to the forcing function given by Eq. (25) can be computed with Eq. (18) which gives $u(s) = \mathcal{H}(s)F(s)$, in which $F(s)$ is the Laplace transform of $F(t)$ given by Eq. (25)

$$F(s) = \int_0^\infty F(t)e^{-st}dt = F_o \int_0^{T_e} e^{-st}dt = F_o \left( \frac{1}{s} - \frac{1}{s}e^{-sT_e} \right) \quad (26)$$

and $\mathcal{H}(s)$ is the complex dynamic flexibility, given by Eq. (18). Accordingly, under a rectangular pulse-load

$$u(s) = F_o \left[ \frac{1}{s} \mathcal{H}(s) - \frac{1}{s} \mathcal{H}(s) e^{-sT_e} \right] \quad (27)$$

### A. Response of the three-parameter inertoelastic solid to a rectangular pulse

Recognizing that $\mathcal{H}(s)/s = \mathcal{C}(s)$ is the complex creep function expressed by Eq. (19) and after expanding the first term in the bracket of the right-hand-side of Equation (18) in partial fractions:

$$\frac{1}{s}\frac{\omega_R^2}{s^2+\omega_R^2} = \frac{1}{s} - \frac{s}{s^2+\omega_R^2} \tag{28}$$

the inverse Laplace transform of Eq. (27) gives

$$u(t) = F_0\left[J(t) - \frac{1}{k_1}\mathcal{L}^{-1}\left(\frac{1}{s}e^{-sT_e} - \frac{s}{s^2+\omega_R^2}e^{-sT_e}\right) - \frac{1}{k_1}\frac{\omega_R^2}{\omega_{R2}^2}\cos[\omega_R(t-T_e)]\right] \tag{29}$$

Upon replacing in Eq. (29) the expression of $J(t)$ given by Equation (20) and after performing the remaining inverse Laplace transforms appearing in Eq. (29), the response of the three-parameter inertoelastic solid shown in Fig. 9 when subjected to the rectangular pulse loading given by Eq. (25) is

$$u(t) = \frac{F_0}{k_1}\left[U(t-0) - U(t-T_e) - \frac{\beta}{1+\beta}\cos(\omega_R t) + \frac{\beta}{1+\beta}[\cos[\omega_R(t-T_e)]]\right] \tag{30}$$

Clearly, when the duration of the acute shock (rectangular pulse) is an integer multiple of a day $\left(T_e = n(24h) = n\frac{24h}{2\pi}2\pi = n\frac{2\pi}{\omega_R}\right)$, the last two cosine terms in Eq. (30) cancel and the displacement response reduces to $u(t) = (F_0/k_1)[U(t-0) - U(t-T_e)]$. On the other hand, when the duration of the rectangular pulse is not an integer multiple of a day (say $T_e = 138\ hours$) the displacement response of a city, $u(t)$, upon subjected to the rectangular pulse exhibits perpetual oscillations (see Fig. 14(c)).

Consider now a city operating under normal conditions, therefore its mean-square displacement $<r^2(t)>$ of its citizens is expressed with Eq. (12). For the simplest admissible mechanical analog shown in Figure 9, the corresponding normalized mean-square displacement $k_1 <r^2(t)>/W = <r^2(t)>/A_o$ given by Eq. (21) is plotted at the top of Fig. 14(a). While the city operates under normal conditions, it is suddenly subjected to an acute shock at some time $t_0 = 0$ that is described with the rectangular pulse given by Eq. (25) with a duration $T_e$ as shown in Fig. 14(b). In the response analysis shown in Fig. 14, we have selected a duration of the pulse excitation to be a non-integer multiple of a day ($T_e = 138\ h = 5\ days + 18\ hours$), so that $u(t)$ as expressed by Eq. (30) exhibits oscillations. The normalized response of the city, $k_1 u(t)/F_0$, to this rectangular pulse alone is offered by Eq. (30) and plotted in Fig. 14(c). Given that in this work we consider the city to be a linear network, the overall response upon the city is subjected to the acute shock (rectangular pulse) is the superposition of its steady-state response under normal conditions given by Eqs. (20) or (21) and shown in Fig. 14(a) (response to a unit step loading at the distant past that produces the creep compliance, $J(t)$) and its response due to the rectangular pulse at time $t_0$ given by Eq. (30) and shown in Fig. 14(c). Accordingly, the "post event" response of a city as is described with the three-parameter inertoelastic solid is

$$\underbrace{k_1 J(t)}_{\text{Response under normal condition: Eq. (20).}} - \underbrace{\frac{k_1}{F_o}u(t)}_{\text{Response due to acute shock. Eq. (30).}} = \overbrace{U(t-T_e) - \frac{\beta}{1+\beta}[\omega_R(t-T_e)]}^{\text{Post-event response}} \tag{31}$$

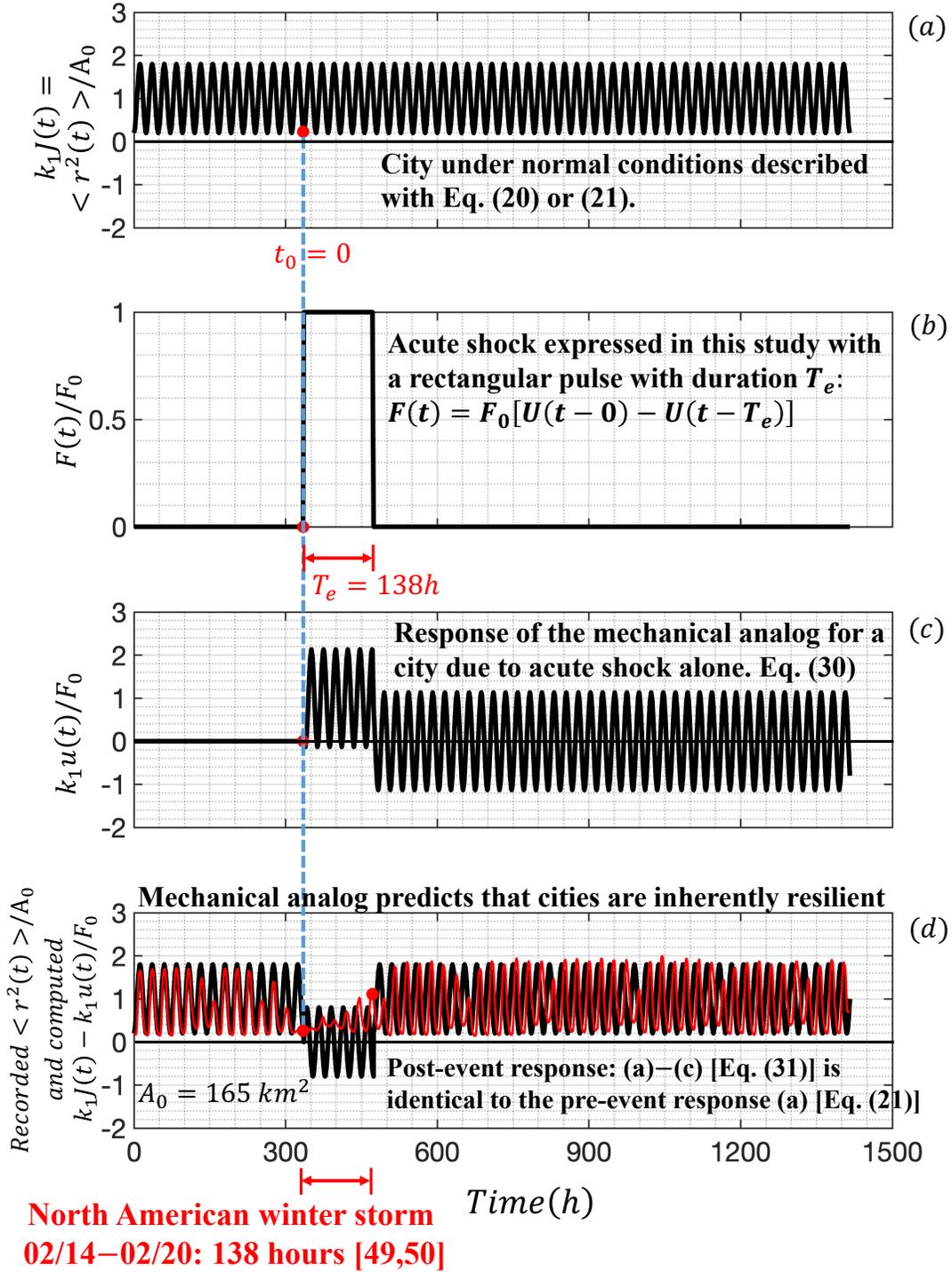

**FIG. 14.** (a) Normalized behavior of a city under normal conditions as given by Eqs. (20) or (21); (b) Acute shock on a city expressed with a rectangular pulse with duration $T_e = 138h$ that is not an integer multiple of a day= $24h$; (c) Normalized response of a city to a rectangular pulse alone with duration $T_e = 138h$; (d) Pre- and post-event response of a city ((a)−(c), solid dark lines) which is identical to the pre-event response of the city (a). The thin red lines is the normalized MSD of the Dallas metroplex prior, during and after the North American winter storm during the $3^{rd}$ week of February 2021 as computed from the recorded GPS location data.

The minus sign (rather than a + sign) in the left-hand side of Eq. (31) is because the acute shock is expected to suppress the normal activity of the city as expressed by Eqs. (20) or (21). The right-hand side of Eq. (31) that results after subtracting Eq. (30) from Eq. (20) is precisely of the same form as Eq. (20) and uncovers the remarkable result that upon the expiration of the acute shock (rectangular pulse), the mechanical model shown Fig. 9 and expressed by Eq. (16) predicts that the city reverts immediately to its normal activity (same $<r^2(t)>$ as before the acute shock), implying that cities when subjected to a rectangular pulse of any duration, $T_e$ are inherently and invariably resilient. Fig. 14(d) plots the pre– and post-event response as predicted by Eq. (31) and compares the analytical prediction to the recorded response of the Dallas metroplex following the 2021 North America cold storm during the $3^{rd}$ week of February 2021 for which we have used a pulse excitation duration $T_e = 138h$.

The remarkable result that cities are inherently resilient to acute shocks as indicated by Eq. (31) is also illustrated in Fig. 15 for the case where the duration of the rectangular shock is an integer multiple of a day ($T_e = 144h$) as shown in Fig. 15(b).

Given the linearity of the candidate constitutive model for a city expressed by Eq. (16) and illustrated in Fig. 7, an alternative way to reach Eq. (30) is through the convolution integral

$$u(t) = \int_{0^-}^{t} h(t-\xi)F(\xi)d\xi \tag{32}$$

where $F(\xi)$ is the time-history of the acute shock—in this analysis the rectangular pulse given by Eq. (25); and $h(t)$ is the impulse response function [47,61-65] that is defined as the resulting displacement history $u(t)$, due to an impulsive force input $F(t) = \delta(t-\xi)$ with $\xi < t$. Given that $\delta(t-\xi) = dU(t-\xi)/dt$, the impulse response function of a linear system $h(t) = dJ(t)/dt$ where $J(t)$ is its creep compliance (displacement history due to a unit step force). Accordingly, the impulse response function of the three-parameter inertoelastic solid described with Eq. (16) and illustrated by Fig. 7 is

$$h(t) = \frac{dJ(t)}{dt} = \frac{1}{k_1}\left[\delta(t-0) + \frac{\beta}{1+\beta}\omega_R \sin(\omega_R t)\right] \tag{33}$$

where $J(t)$ is given by Eq. (20) and $\delta(t-0)$ is the Dirac delta function [57].
Substitution of the expression of the impulse response function given by Eq. (33) into Eq. (32) after using the forcing function $F(\xi)$ given by Eq. (25) yields

$$u(t) = \frac{F_0}{k_1}\left[\int_{0^-}^{t}\delta(t-\xi)[U(\xi-0)-U(\xi-T_e)]d\xi + \frac{\beta}{1+\beta}\omega_R\int_{0}^{T_e}\sin[\omega_R(t-\xi)]d\xi\right] \tag{34}$$

Using the symmetry of the Dirac delta function $\delta(t-\xi) = \delta(\xi-t)$ [57], the first integral appearing within the brackets of Eq. (34) gives

$$\int_{0^-}^{t}\delta(\xi-t)[U(\xi-0)-U(\xi-T_e)]d\xi = U(t-0)-U(t-T_e) \tag{35}$$

and after expanding the sine function in the second integral, Eq. (34) assume the form

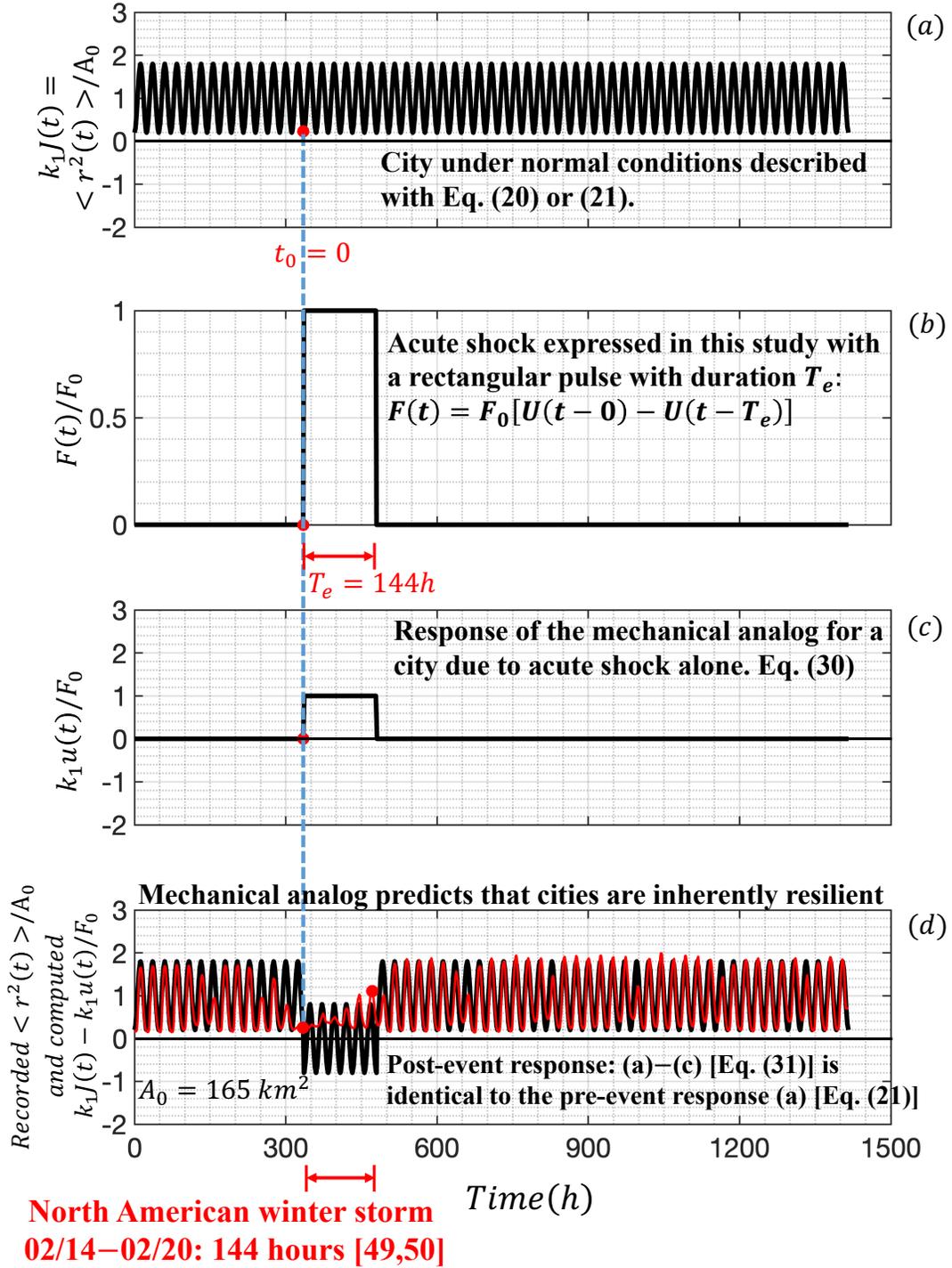

**FIG. 15.** (a) Normalized behavior of a city under normal conditions as given by Eqs. (20) or (21); (b) Acute shock on a city expressed with a rectangular pulse with duration $T_e = 144h$ that is an integer multiple of a day= $24h$; (c) Normalized response of a city to a rectangular pulse alone with duration $T_e = 144h$; (d) Pre- and post-event response of a city ((a)−(c), solid dark lines) which is identical to the pre-event response of the city (a). The thin red lines is the normalized MSD of the Dallas metroplex prior, during and after the North American winter storm during the $3^{rd}$ week of February 2021 as computed from the recorded GPS location data.

$$u(t) = \frac{F_0}{k_1}\left[[U(t-0) - U(t-T_e)] + \frac{\beta}{1+\beta}\omega_R\left[\sin(\omega_R t)\int_0^{T_e}\cos(\omega_R \xi)\,d\xi - \cos(\omega_R t)\int_0^{T_e}\sin(\omega_R \xi)\,d\xi\right]\right] \quad (36)$$

Upon evaluating the cosine and sine integrals in the right-hand side of Eq. (36), the displacement response history offered by Eq. (30) is recovered.

This second approach, where the displacement response is evaluated with the convolution integral given by Eq.(32), is employed to calculate the response from the amplitude-modulation model for which we do not have a complex frequency response function, $\mathcal{H}(s)$ at this time.

### B. Response of the amplitude-modulation model to a rectangular pulse

For a city that the mean-square displacement of its citizens $\langle r^2(t)\rangle$ is best described with the amplitude-modulation model expressed with Eq. (24) the corresponding creep compliance is

$$J(t) = \frac{\langle r^2(t)\rangle}{A_0 k} = \frac{1}{k}\left\{U(t-0) - C\left[\cos(\omega_R t) + \frac{\gamma}{2}\left(\sin\left(\frac{8}{7}\omega_R t\right) - \sin\left(\frac{6}{7}\omega_R t\right)\right)\right]\right\} \quad (37)$$

and its impulse response function is

$$h(t) = \frac{dJ(t)}{dt} = \frac{1}{k}\left\{\delta(t-0) + C\omega_R\left[\sin(\omega_R t) - \frac{\gamma}{2}\left(\frac{8}{7}\cos\left(\frac{8}{7}\omega_R t\right) - \frac{6}{7}\cos\left(\frac{6}{7}\omega_R t\right)\right)\right]\right\} \quad (38)$$

Substitution of the expression of the impulse response function given by Eq. (38) into Eq. (32) after using that the forcing function $F(\xi)$ is given by Eq. (25), the normalized displacement response from the amplitude-modulation model when subjected to a rectangular pulse of duration $T_e$ is

$$\frac{k}{F_0}u(t) = U(t-0) - U(t-t_e) + C\left[-\cos(\omega_R t) + \cos(\omega_R(t-T_e))\right. \\ \left. - \frac{\gamma}{2}\left(\sin\left(\frac{8}{7}\omega_R t\right) - \sin\left(\frac{8}{7}\omega_R(t-T_e)\right) - \sin\left(\frac{6}{7}\omega_R t\right) + \sin\left(\frac{6}{7}\omega_R(t-T_e)\right)\right)\right] \quad (39)$$

Again, when the duration of the rectangular pulse is not an integer multiple of a day (say $T_e = 138\ hours$), the displacement response of a city $u(t)$ described with the amplitude modulation model exhibits perpetual oscillations as shown in Fig. 16(c).

Consider now a city operating under normal conditions and the mean-square displacement $\langle r^2(t)\rangle$ of its citizens is best described by the amplitude-modulation model given by Eq. (24) and plotted at the top of Fig. 16(a). While the city operates under normal conditions, at some time $t_0 = 0$ is suddenly subjected to an acute shock that is described with a rectangular pulse with duration $T_e$ as shown in Fig. 16(b). In the response analysis shown in Fig. 16 we have selected a duration of the pulse−excitation to be a non-integer multiple of a day ($T_e = 138h = 5\ days + 18\ hours$) so that the city response $u(t)$ as expressed by Eq. (39) exhibits oscillations. The overall response upon the city is subjected to the acute shock (post-event response) is the superposition of its steady-state response under normal conditions given by Eq. (24) and shown in Fig. 16(a) and its response due to the rectangular pulse given by Eq. (39) and shown in Fig. 16(c). Accordingly, the "post-event" response of a city described with the amplitude-modulation model is

$$\underbrace{kJ(t)}_{} - \underbrace{\frac{k}{F_0}u(t)}_{} = \overbrace{U(t-T_e) - C\left[\cos(\omega_R(t-T_e)) + \frac{\gamma}{2}\left(\sin\left(\frac{8}{7}\omega_R(t-T_e)\right) - \sin\left(\frac{6}{7}\omega_R(t-T_e)\right)\right)\right]}^{\text{Post-event response}} \quad (40)$$

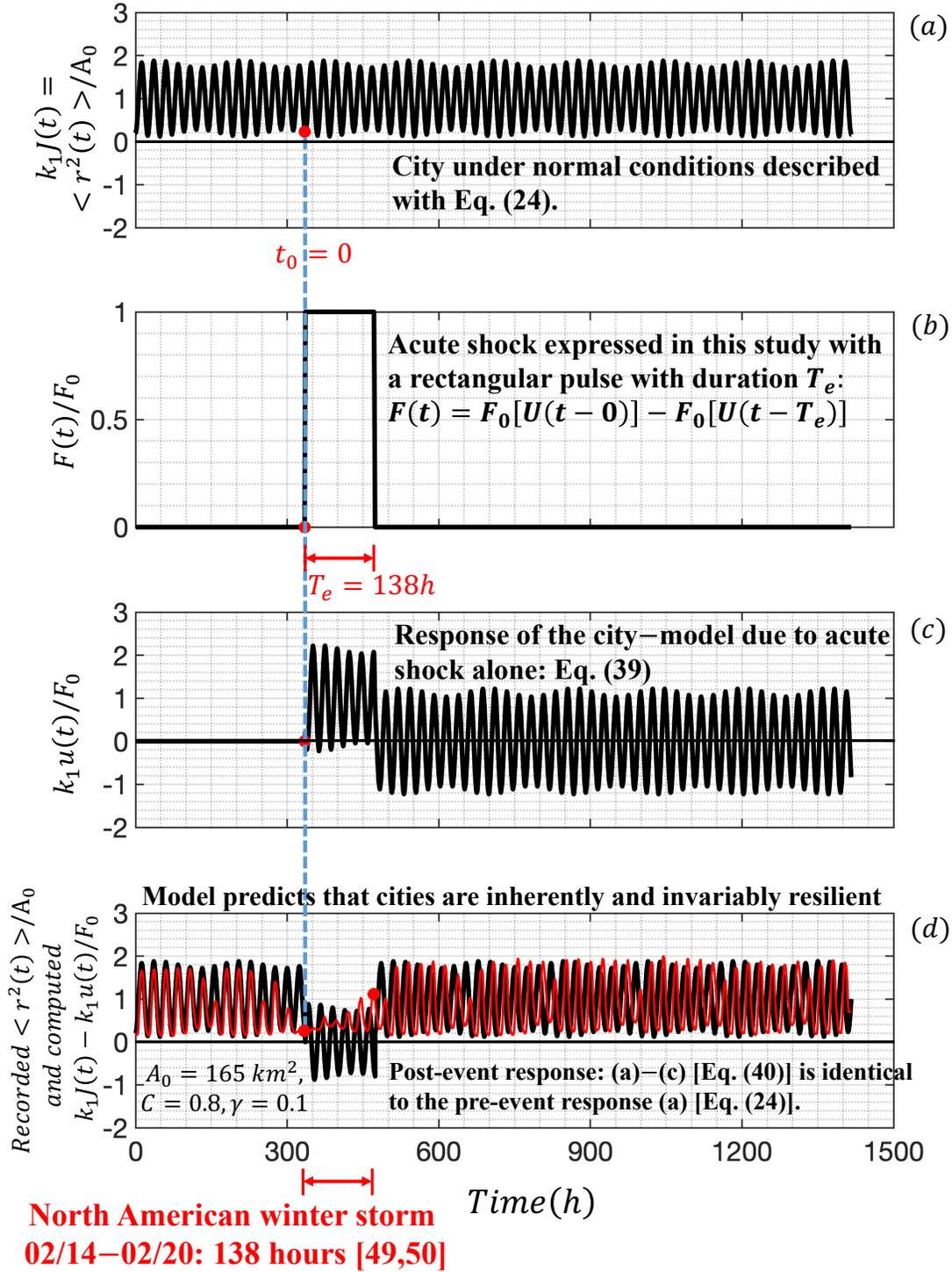

**FIG. 16.** (a) Normalized behavior of a city under normal conditions as given by Eqs. (24); (b) Acute shock on a city expressed with a rectangular pulse with duration $T_e = 138h$ that is not an integer multiple of a day= $24h$; (c) Normalized response of a city to a rectangular pulse alone with duration $T_e = 138h$; (d) Pre- and post-event response of a city ((a)−(c), solid dark lines) which is identical to the pre-event response of the city (a). The thin red lines is the normalized MSD of the Dallas metroplex prior, during and after the North American winter storm during the $3^{rd}$ week of February 2021 as computed from the recorded GPS location data.

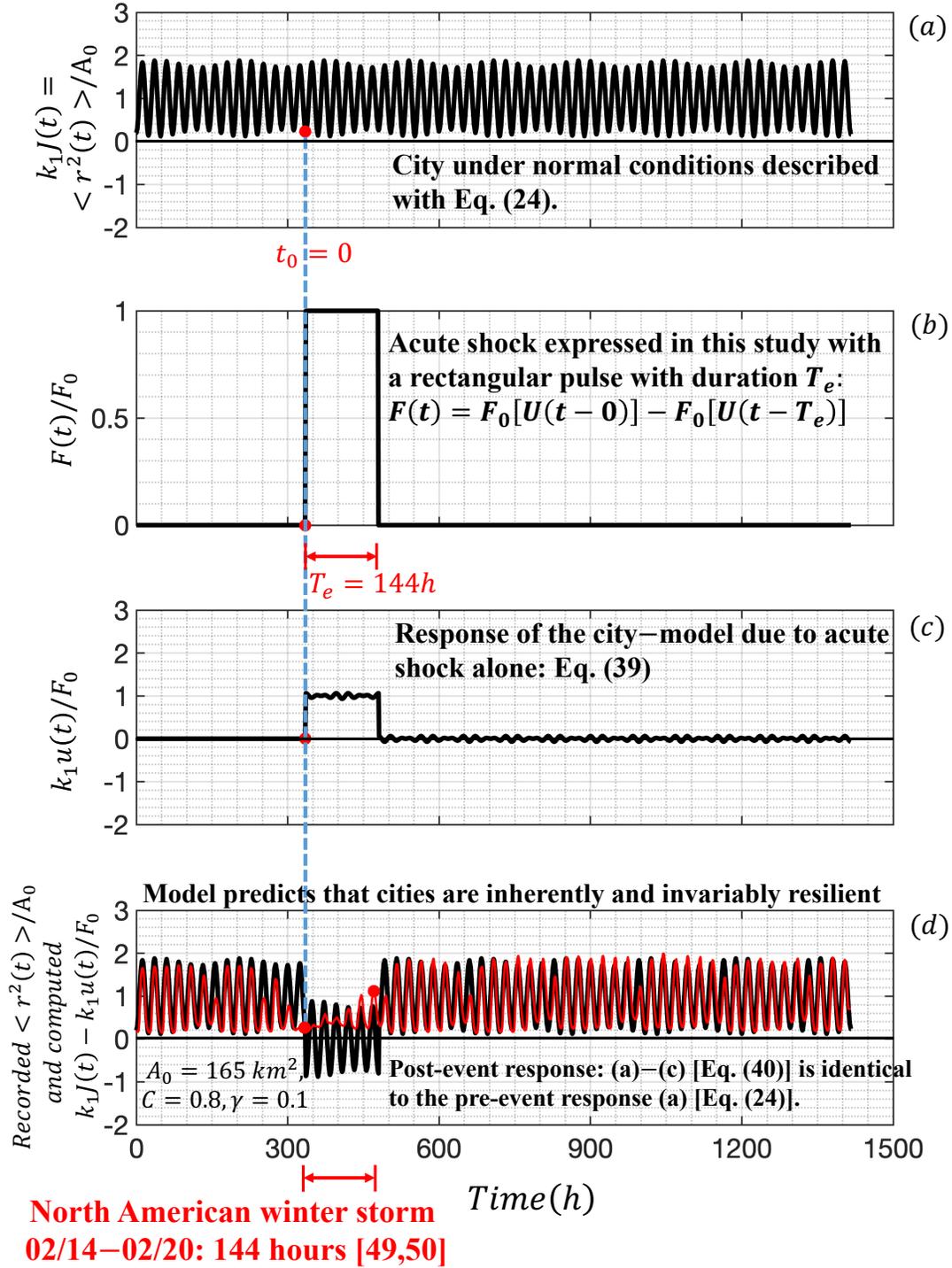

**FIG. 17.** (a) Normalized behavior of a city under normal conditions as given by Eqs. (24); (b) Acute shock on a city expressed with a rectangular pulse with duration $T_e = 144h$ that is an integer multiple of a day= $24h$; (c) Normalized response of a city to a rectangular pulse alone with duration $T_e = 144h$; (d) Pre- and post-event response of a city ((a)−(c), solid dark lines) which is identical to the pre-event response of the city (a). The thin red lines is the normalized MSD of the Dallas metroplex prior, during and after the North American winter storm during the $3^{rd}$ week of February 2021 as computed from the recorded GPS location data.

↳ Response due to acute shock: Eq. (39)

↳ Response under normal conditions Eq. (24).

The right-hand side of Eq. (40), that results after subtracting Eq. (39) from Eq. (24), is precisely of the same form as Eq. (24) and confirms our previous finding that, regardless of the sophistication of the linear mathematical model used to approximate the mean-square displacement of the citizens of a city, upon the expiration of the acute shock (rectangular pulse), the city reverts immediately to its normal pre-event activities (same $\langle r^2(t) \rangle$ as before the acute shock) implying that cities when subjected to a rectangular pulse are inherently resilient within the context of *engineering resilience* [33].

The remarkable result—that cities are inherently resilient to acute shocks, regardless the sophistication of the linear solid-like model that approximates the recorded mean-square displacement is also illustrated in Fig. 17 for the case where the duration of the rectangular pulse is an integer multiple of a day as shown in Fig. 17(b). For the amplitude-modulation model, even when the duration of the rectangular shock is an integer number of days ($T_e = 144h = 6\ days$) the sine functions appearing in Eq. (39) do not cancel because of the multiplication factors 8/7 and 6/7, resulting to the small undulations that override the rectangular pulse shown in Fig. 17(c). The predictions of our mathematical models are in remarkable good agreement with the recorded response of the Dallas metroplex following the 2021 North American winter storm during which over 4 million people lost power due to a resulting state-wide power crisis in Texas causing unforeseen damage to its power grid [49,50].

## VI.  SUMMARY

Following the ever-growing number of qualitative studies on urban resilience [10] in association with current trends for developing a quantitative, science-based, predictive framework for the behavior/response analysis of cities [2,6]; in this work we employ concepts from statistical mechanics and microrheology to develop a mechanical analog for cities. Our study concentrates on a *single-state equilibrium* as defined by [33] which is the capacity of a system to revert to its post-disturbance equilibrium state—also known as *engineering resilience* and hinges upon the premise that a dependable indicator of the engineering resilience of a city is whether the average mobility pattern of its citizens following an acute shock matches the average mobility pattern before the shock.

We built on the remarkable success of microrheology [22-32] in which the macroscopic frequency and time response functions of complex viscoelastic materials are extracted by monitoring the thermally driven Brownian motion of probe microparticles suspended within the viscoelastic material and we computed the mean-square displacements (ensemble averages) of thousands of cell-phone users from GPS location data to establish the creep compliance and the resulting impulse response function of a city. The derivation of these time-response functions allows for the synthesis of simple solid-like mechanical analogs with increasing sophistication which are shown to have engineering significance.

Arguments from physical modeling led to the synthesis of the simplest mechanical analog for a city which is the three-parameter inertoelastic solid which is a parallel connection of an elastic spring with an inertoelastic fluid (a spring-inerter in series connection). The three-parameter inertoelastic solid yields a creep compliance which is essentially an elevated cosine that captures the daily oscillations of the mean-square displacement of the citizens of a city in association that a city "never sleeps".

Recorded GPS location data from the Dallas metroplex and the San Francisco Bay Area indicate that in addition to the daily oscillations, the mean-square displacement of cities exhibits a weekly amplitude-modulation with people moving more during the weekdays and less on Sundays; while, less people are staying at home during the weekends than during the weekdays. These observations motivated the synthesis and response analysis of a solid-like, amplitude-modulation model which captures satisfactorily both the daily and weekly periodicity. Given their linear structure, both the three-parameter inertoelastic model and the more sophisticated amplitude-modulation solid-like model predict, that upon a city is subjected to an acute shock that was modeled with a rectangular pulse with arbitrary finite duration, it recovers immediately its initial state (pre-event response), suggesting that cities are inherently resilient within the context of engineering resilience [33]. The mathematical predictions from the mechanical analogs developed in this study confirm similar findings by other investigators, reached with different analysis methodologies, that cites are remarkably robust [6].

## ACKNOWLEDGEMENTS

This work was partially funded by the SMU Research Clusters on "Using New Data Sources", by the Hunt Institute for Engineering and Humanity; and by the SMU-Lyle interdisciplinary seed funding initiative. Data analysis was performed on the ManeFrame II (M2): high-performance computing (HPC) cluster supported by the SMU Research and Data Science Services.

## REFERENCES

[1] J. Gao, S. V. Buldyrev, H. E. Stanley, and S. Havlin, Nat. phys. **8**, 40 (2012).
[2] L. M. Bettencourt, Science **340**, 1438 (2013).
[3] S. Lhomme, D. Serre, Y. Diab, and R. Laganier, *Urban technical networks resilience assessment* (CRC Press London, 2013), pp. 109-117.
[4] S. S. Cruz, J. P. T. Costa, S. Á d. Sousa, and P. Pinho, *Resilience and spatial dynamics* (Springer, 2013), pp. 53-69.
[5] M. Brockerhoff, Population and Development Rev. **24**, 883 (1998).
[6] L. Bettencourt and G. West, Nature **467**, 912 (2010).
[7] D. R. Godschalk, Nat. Hazards Rev. **4**, 136 (2003).
[8] J. Da Silva, S. Kernaghan, and A. Luque, J. Urban Sustainable Development **4**, 125 (2012).
[9] Y. Jabareen, Cities **31**, 220 (2013).
[10] S. Meerow, J. P. Newell, and M. Stults, Landscape and Urban Planning. **147**, 38 (2016).
[11] L. M. Bettencourt and J. Kaur, Proceeding of the National Academy of Sciences **108**, 19540 (2011).
[12] L. M. Bettencourt, J. Lobo, D. Strumsky, and G. B. West, Plos One **5**, e13541 (2010).
[13] E. N. Bacon and K. Walduck, *Design of cities* (Thames and Hudson, London, United Kingdom, 1974).
[14] G. B. West, J. H. Brown, and B. J. Enquist, Science **276**, 122 (1997).

[15] G. B. West, J. H. Brown, and B. J. Enquist, Science **284**, 1677 (1999).

[16] H. Samaniego and M. E. Moses, JTLU **1**, 21 (2008).

[17] J. S. Waters, C. Tate Holbrook, J. H. Fewell, and J. F. Harrison, The American Naturalist **176**, 501 (2010).

[18] D. S. Dendrinos and H. Mullally, *Urban evolution: studies in the mathematical ecology of cities* (Oxford University Press, USA, 1985).

[19] M. C. Gonzalez, C. A. Hidalgo, and A. Barabasi, Nature **453**, 779 (2008).

[20] A. Noulas, S. Scellato, R. Lambiotte, M. Pontil, and C. Mascolo, Plos One **7** (2012).

[21] D. Brockmann, L. Hufnagel, and T. Geisel, Nature **439**, 462 (2006).

[22] T. G. Mason and D. A. Weitz, Phys. Rev. Lett. **74**, 1250 (1995).

[23] T. G. Mason, H. Gang, and D. A. Weitz, J. Opt. Soc. Am. **14**, 139 (1997).

[24] T. G. Mason, K. Ganesan, J. H. van Zanten, D. Wirtz, and S. C. Kuo, Phys. Rev. Lett. **79**, 3282 (1997).

[25] F. Gittes, B. Schnurr, P. D. Olmsted, F. C. MacKintosh, and C. F. Schmidt, Phys. Rev. Lett. **79**, 3286 (1997).

[26] B. Schnurr, F. Gittes, F. C. MacKintosh, and C. F. Schmidt, Macromolecules **30**, 7781 (1997).

[27] T. G. Mason, Rheol. Acta **39**, 371 (2000).

[28] T. A. Waigh, Rep. Prog. Phys. **68**, 685 (2005).

[29] T. Li, S. Kheifets, D. Medellin, and M. G. Raizen, Science **328**, 1673 (2010).

[30] R. Huang, I. Chavez, K. M. Taute, B. Lukić, S. Jeney, M. G. Raizen, and E. Florin, Nat. Phys. **7**, 576 (2011).

[31] T. Li and M. G. Raizen, Ann. Phys. **525**, 281 (2013).

[32] A. Jannasch, M. Mahamdeh, and E. Schäffer, Phys. Rev. Lett. **107**, 228301 (2011).

[33] C. S. Holling, *Engineering resilience versus ecological resilience*, 1996), Vol. 31, p. 32.

[34] T. J. Campanella, JAPA **72**, 141 (2006).

[35] W. A. H. Hamilton, in *Proceedings of the Institution of Civil Engineers - Urban Design and Planning, 2009* (ICE Publishing, 2009), pp. 109-121.

[36] J. E. Lamond and D. G. Proverbs, (Thomas Telford Ltd, 2009) pp. 63–70.

[37] *Equirectangular approximation*. https://www.movable-type.co.uk/scripts/latlong.html

[38] J. S. Bendat and A. G. Piersol, *Random data: analysis and measurement procedures* (John Wiley & Sons, 2011).

[39] S. H. Crandall and W. D. Mark, *Random vibration in mechanical systems* (1963).

[40] A. Einstein, Ann. Phys. **4**, 549 (1905).

[41] L. D. Landau and E. M. Lifshitz, *Fluid mechanics. Translated from the Russian by JB Sykes and WH Reid*, Vol. 6 (1987).

[42] G. E. Uhlenbeck and L. S. Ornstein, Phys. Rev. **36**, 823 (1930).

[43] V. S. Volkov and G. V. Vinogradov, J. Non-Newtonian Fluid. Mech. **15**, 29 (1984).

[44] R. F. Rodriguez and E. Salinas-Rodriguez, J. uPhys. A **21**, 2121 (1988).

[45] N. Makris, Phys. Rev. E **101**, 052139 (2020).

[46] M. C. Wang and G. E. Uhlenbeck, Rev. Mod. Phys. **17**, 323 (1945).

[47] N. Makris, Phys. Fluids. **33**, 072014 (2021).

[48] J. Boussinesq, CR Acad. Sci. **100**, 935 (1885).

[49] *North American winter storm*. https://www.weather.gov/fwd/Feb-2021-WinterEvent

[50] *February 2021 North American winter storm,* https://www.noaa.gov/news/us-had-its-coldest-february-in-more-than-30-years

[51] N. Makris, J. Eng. Mech. **143**, 04017123 (2017).

[52] N. Makris, Meccanica **53**, 2237 (2018).

[53] A. Palmer, J. Xu, and D. Wirtz, Rheol. Acta. **37**, 97 (1998).

[54] J. Xu, V. Viasnoff, and D. Wirtz, Rheol. Acta. **37**, 387 (1998).

[55] R. Evans, M. Tassieri, D. Auhl, and T. A. Waigh, Phys. Rev. E **80**, 012501 (2009).

[56] N. Makris, Meccanica **54**, 19 (2019).

[57] M. J. Lighthill, *An introduction to Fourier analysis and generalised functions* (Cambridge University Press, 1958).

[58] L. Gray and R. Graham, *Radio transmitters* (McGraw-Hill New York, Toronto, London, 1961).

[59] H. W. Silver, *The ARRL handbook: for radio communications* (AARL, 2011).

[60] S. Abbar, T. Zanouda, and J. Borge-Holthoefer, UrbComp'16, San Francisco, USA. (2016).

[61] R. W. Clough and J. Penzien, *Dynamics of structures* (McGraw-Hill, New York, NY, 1970).

[62] C. M. Harris and C. E. Crede, *Shock and vibration handbook* (McGraw-Hill, New York, NY, USA, 1976).

[63] A. V. Oppenheim and R. W. Schafer, *Digital signal processing* (Prentice-Hall, Inc., Englewood Cliffs, NJ, 1975), pp. 137-148.

[64] J. G. Reid, *Linear system fundamentals: Continuous and discrete, classic and modern* (McGraw-Hill Science, Engineering & Mathematics, 1983).

[65] N. Makris, Soft Matter **17**, 5410 (2021).
28